\documentclass[12pt]{article}
\usepackage[dvips]{graphicx}
\usepackage{epsfig,cite}
\usepackage {amsmath}
\usepackage{color}
\usepackage{amssymb}
\usepackage{slashed}
\include{epsf}
\newcount\timecount
\newcount\hours \newcount\minutes  \newcount\temp \newcount\pmhours
\hours = \time
\divide\hours by 60
\temp = \hours
\multiply\temp by 60
\minutes = \time
\advance\minutes by -\temp
\def\hour{\the\hours}
\def\minute{\ifnum\minutes<10 0\the\minutes
            \else\the\minutes\fi}
\def\clock{
\ifnum\hours=0 12:\minute\ AM
\else\ifnum\hours<12 \hour:\minute\ AM
      \else\ifnum\hours=12 12:\minute\ PM
            \else\ifnum\hours>12
                 \pmhours=\hours
                 \advance\pmhours by -12
                 \the\pmhours:\minute\ PM
                 \fi
            \fi
      \fi
\fi
}

\def\monthname{\relax\ifcase\month 0/\or January\or February\or
   March\or April\or May\or June\or July\or August\or September\or
   October\or November\or December\else\number\month/\fi}

\def\bold#1{\setbox0=\hbox{$#1$}%
     \kern-.025em\copy0\kern-\wd0
     \kern.05em\copy0\kern-\wd0
     \kern-.025em\raise.0433em\box0 }

\def\beq{\begin{equation}}
\def\eeq{\end{equation}}

\def\ga{\mathrel{\raise.3ex\hbox{$>$\kern-.75em\lower1ex\hbox{$\sim$}}}}
\def\la{\mathrel{\raise.3ex\hbox{$<$\kern-.75em\lower1ex\hbox{$\sim$}}}}
\def\gev{{\rm \, Ge\kern-0.125em V}}
\def\tev{{\rm \, Te\kern-0.125em V}}
\def\gyr{{\rm \, G\kern-0.125em yr}}

\def\tbt{\tan \beta}
\def\tanb{\tbt}

\def\gappeq{\mathrel{\rlap {\raise.5ex\hbox{$>$}}
{\lower.5ex\hbox{$\sim$}}}}
\def\lappeq{\mathrel{\rlap{\raise.5ex\hbox{$<$}}
{\lower.5ex\hbox{$\sim$}}}}
\def\Toprel#1\over#2{\mathrel{\mathop{#2}\limits^{#1}}}

\def\m12{m_{1\!/2}}

\def\bea{\begin{eqnarray}}
\def\eea{\end{eqnarray}}

\def\beqar{\begin{eqnarray}}
\def\eeqar{\end{eqnarray}}

\begin{document}

\begin{titlepage}
\pagestyle{empty}
\baselineskip=21pt
\rightline{CERN-PH-TH/2011-130, KCL-PH-TH/2011-13, LCTS/2011-02}
\rightline{UMN--TH--3004/11, FTPI--MINN--11/14}
\vskip 0.1in
\begin{center}
{\large {\bf Galactic-Centre Gamma Rays in CMSSM Dark Matter Scenarios}}

\end{center}
\begin{center}
 {\bf John~Ellis}$^{1,2}$,
{\bf Keith~A.~Olive}$^{3}$
and {\bf Vassilis~C.~Spanos}$^4$
\vskip 0.1in
{\small {\it
$^1${TH Division, Physics Department, CERN, CH-1211 Geneva 23, Switzerland}\\
$^2${Theoretical Particle Physics and Cosmology Group, Physics Department, \\
King's College London, London WC2R 2LS, UK}\\
$^3${William I. Fine Theoretical Physics Institute \& Department of Physics, \\
University of Minnesota, Minneapolis, MN 55455, USA}\\
$^4${Institute of Nuclear Physics, NCSR ``Demokritos'', GR-15310 Athens, Greece}} \\
}
\vskip 0.2in
{\bf Abstract}
\end{center}
\baselineskip=18pt \noindent
{\small

We study the production of $\gamma$ rays via LSP annihilations in the core of the Galaxy
as a possible experimental signature of the constrained minimal supersymmetric extension of the Standard
Model (CMSSM), in which supersymmetry-breaking parameters are assumed to be universal at the
GUT scale, assuming also that the LSP is the lightest neutralino $\chi$.
The part of the CMSSM parameter space that is compatible with the measured astrophysical
density of cold dark matter is known to include a ${\tilde \tau_1} - \chi$ coannihilation strip, a
focus-point strip where $\chi$ has an enhanced Higgsino component, and a funnel at large $\tan \beta$ where the
annihilation rate is enhanced by the poles of nearby heavy MSSM Higgs bosons, $A/H$. We calculate
the total annihilation rates, the fractions of annihilations into different Standard Model
final states and the resulting fluxes of $\gamma$ rays for CMSSM scenarios along these strips. 
We observe that typical annihilation rates
are much smaller in the coannihilation strip for $\tan \beta = 10$ than along the focus-point strip
or for $\tan \beta = 55$, and that the annihilation
branching ratios differ greatly between the different dark matter strips.
Whereas the current Fermi-LAT data are not sensitive to any of the CMSSM scenarios studied, and 
the calculated $\gamma$-ray
fluxes are probably unobservably low along the coannihilation strip for $\tan \beta = 10$, we find that 
substantial portions of the
focus-point strips and rapid-annihilation funnel regions could be pressured by several more years
of Fermi-LAT data, if understanding of the astrophysical background and/or systematic uncertainties
can be improved in parallel.}


\vfill
\leftline{
June 2011}
\end{titlepage}
\baselineskip=18pt

\section{Introduction}

Relatively soon after the realization that the lightest supersymmetric particle
(LSP),  in particular the lightest neutralino $\chi$, 
could naturally provide the astrophysical dark matter in models in which
$R$-parity is conserved~\cite{EHNOS}, it was suggested that LSP dark
matter annihilations might be detectable via features in the $\gamma$-ray
spectrum \cite{oldgamma}. 
This is now a hot theoretical \cite{newgamma,Bringmann:2007nk,Bergstrom:1997fj,
center,smbh,dwarf,reglast,Bernal:2010ip,cmssmgamma} and
experimental topic, with several experiments~\cite{dgelat,fermi,rpslat,igrblat,Vitale:2009hr,veritas,magic,hess}
studying the cosmic-ray $\gamma$ spectra from a variety of
astrophysical sources such as the galactic centre and bulge as well as
dwarf galaxies. There have been claims of deviations from
calculations of conventional $\gamma$-ray backgrounds, and corresponding claims of
evidence for new physics invoking tailor-made supersymmetric scenarios \cite{hooper}.
In parallel, particularly in the context of searches with accelerator experiments,
there have been many studies of simplified supersymmetric scenarios.
Foremost among such scenarios is the constrained minimal supersymmetric
extension of the Standard Model (CMSSM), in which the soft supersymmetry-breaking
parameters $m_0$, $m_{1/2}$ and $A_0$ are assumed to be universal at the
supersymmetric GUT scale \cite{funnel,cmssm,efgosi,cmssmnew,eoss,cmssmmap}. 

Accelerator experiments and astrophysical searches have
complementary roles to play in elucidating the nature of any dark matter
particles. Accelerator experiments cannot determine whether any
candidate dark matter particle is in fact stable, rather than merely living
long enough to escape from the apparatus, while astroparticle experiments
are limited in their capabilities to disentangle the dynamics of dark matter
models and thereby, for example, verify that they yield the appropriate
cosmological dark matter density. Connecting the accelerator and
astroparticle experiments requires interpreting them within a common
model framework, and in this paper we choose to compare their
sensitivities within the framework of the CMSSM, assuming that $R$-parity is
conserved and that the LSP is the lightest neutralino $\chi$.

The parameters of the CMSSM are the universal scalar mass $m_0$,
gaugino mass $m_{1/2}$ and trilinear supersymmetry-breaking parameter
$A_0$, the ratio of Higgs vevs $\tan \beta$ and the sign of the Higgs mixing
parameter $\mu$. Here we focus on $\mu > 0$, motivated by $g_\mu - 2$ \cite{newBNL,g-2}
and (to a lesser extent) by $b \to s \gamma$ \cite{bsgex}. We study the cases $\tan \beta
=10, 55$, which bracket the phenomenologically-plausible range.
Currently, there is little experimental sensitivity to $A_0$, and we limit our
discussion here to the case $A_0 = 0$.

\begin{figure}[htb]
\begin{center}
\epsfig{file=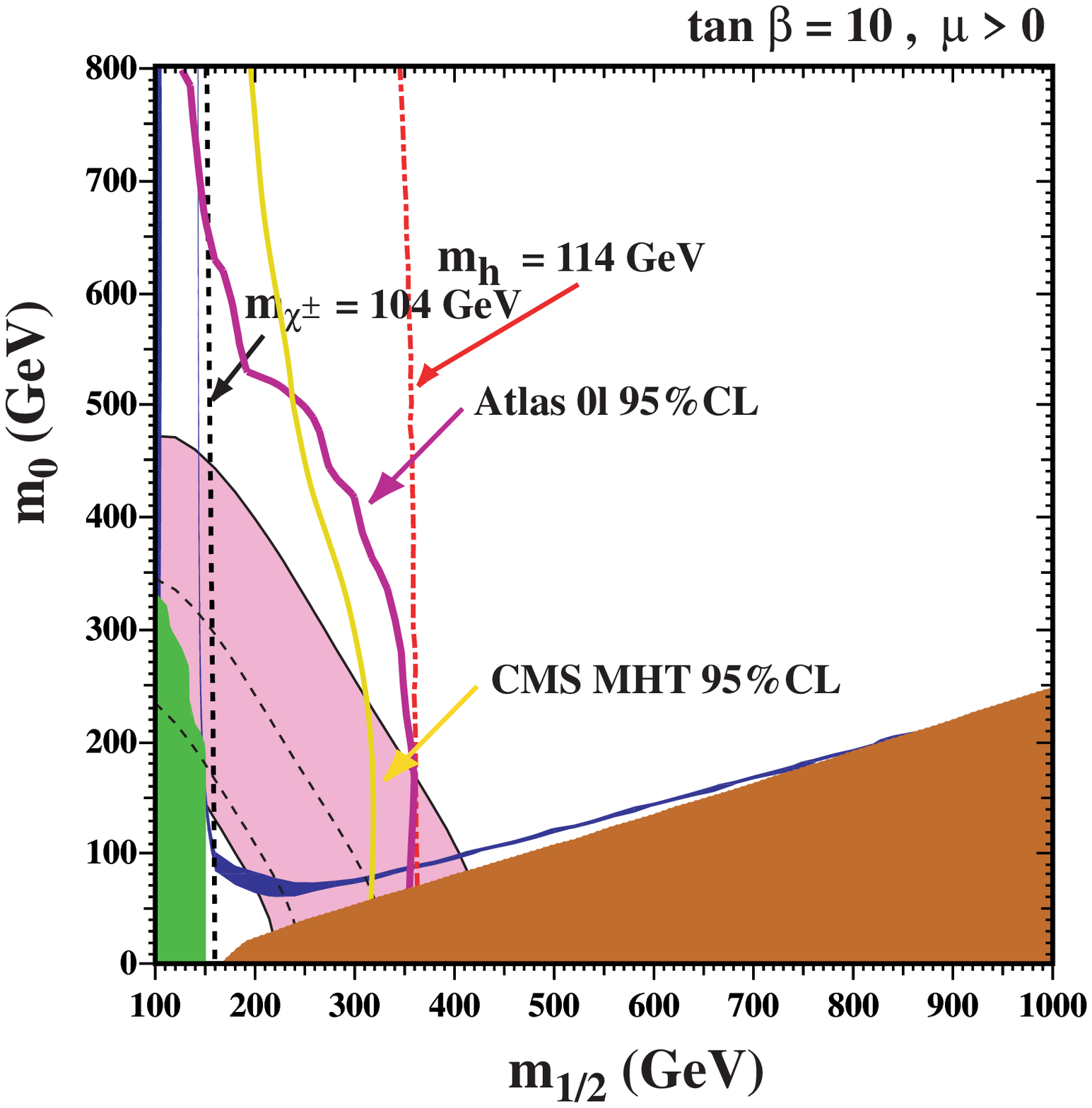,width=0.475\textwidth}
\epsfig{file=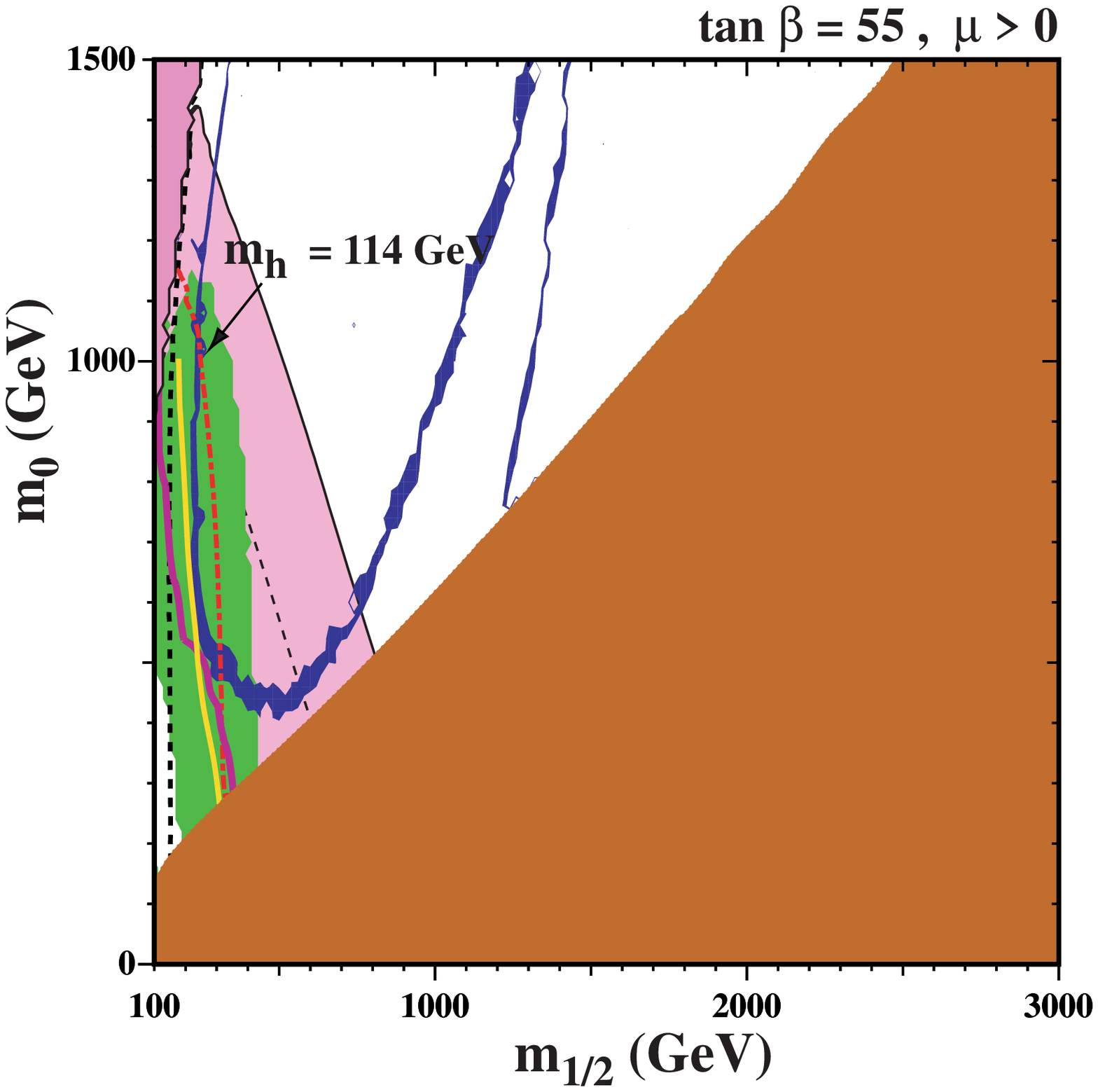,width=0.475\textwidth}
\caption{
\it The $(m_{1/2}, m_0)$ planes in the CMSSM for $\tanb = 10$ (left) and $\tanb=55$ (right),
assuming $\mu > 0$ and $A_0 = 0$, showing the 95\% CL limits imposed by ATLAS and
CMS data (purple and yellow lines, respectively). The regions where the relic LSP
density falls within the range allowed by WMAP and other cosmological observations
appear as strips shaded dark blue. The constraints due to the
absences of charginos and the Higgs boson at LEP are also shown, 
as black dashed and red dot-dashed lines,
respectively. Regions excluded by the requirements of electroweak (EW) symmetry breaking
and a neutral LSP are shaded dark pink and brown, respectively. The green region is
excluded by $b \to s \gamma$, and the pink region is favoured by the supersymmetric
interpretation of the discrepancy between the Standard Model calculation and the
experimental measurement of $g_\mu - 2$ within $\pm 1$ and $\pm 2$ standard
deviations (dashed and solid lines, respectively).}
\label{fig:planes}
\end{center}
\end{figure}

As is well known, the regions of the CMSSM parameter space in which
the dark matter density falls within the narrow range permitted by WMAP and other
cosmological observations \cite{WMAP} may be represented as relatively narrow strips 
in $(m_{1/2}, m_0)$ planes for fixed values of $\tan \beta$ \cite{eoss,bench}, which are
illustrated in Fig.~\ref{fig:planes}. As seen in the left panel, if $\tan \beta = 10$
there is one WMAP strip with $m_0 \ll m_{1/2}$ where the relic density is
reduced into the allowed range by coannihilations of the relic neutralinos $\chi$
with sleptons ${\tilde \ell}$ \cite{stauco}, and another strip at relatively large $m_0$ in the
so-called `focus-point' region \cite{fp} where the relic density is brought into the allowed range 
by enhanced annihilation due to a relatively large Higgsino component in the
composition of the lightest neutralino $\chi$. If $\tan \beta = 55$, as seen in the
right panel of Fig.~\ref{fig:planes}, the
coannihilation strip segues into a funnel region \cite{funnel,efgosi} where the relic density is
brought into the allowed range by rapid annihilations through direct-channel
heavy Higgs resonances $H/A$. Also visible in Fig.~\ref{fig:planes} are the
constraints imposed by $b \to s \gamma$, the LEP lower limits on chargino
and Higgs masses \cite{LEPsusy}, and the region favoured by $g_\mu - 2$ if the apparent
experimental discrepancy with the Standard Model calculation is ascribed
to supersymmetry. Fig.~\ref{fig:planes} also shows the constraints implied by the absence
of any supersymmetric signal in the 2010 LHC data~\cite{lhc}.

In this paper we study the likely sensitivity of searches for $\gamma$ rays from
the galactic centre to LSP annihilations in the CMSSM, taking as examples the
$(m_{1/2}, m_0)$ planes shown in Fig.~\ref{fig:planes} and focusing on the
dark blue WMAP strips, in particular. The first step is to understand relevant
features of CMSSM dark matter annihilation processes, which we study in
Section~2. As we discuss there, the $\chi - \chi$ annihilation cross section is
much smaller in the coannihilation region of the CMSSM for $\tan \beta = 10$
than it is in the focus-point region, or for $\tan \beta = 55$. This is easily
understood, because in the $\tan \beta = 10$ coannihilation strip the relic
density is brought down into the WMAP range by the sum over all 
$\chi - {\tilde \ell}$ coannihilation processes, relative to which $\chi - \chi$ 
annihilation is numerically small. Moreover, only S-wave annihilation is
important in the Universe today, whereas P-wave annihilation also played a role
in the early Universe, particularly for the parameter space
associated with the co-annihilation strip. 
The smallness of the S-wave $\chi - \chi$ annihilation cross section in the
$\tan \beta = 10$ coannihilation strip implies that all astroparticle searches for
dark matter annihilation products, whether they be photons, neutrinos, positrons,
antiprotons or antideuterons, will be relatively insensitive in this region of the
CMSSM parameter space.

A second issue we study in Section~2 is that of the branching fractions for
annihilations into different final Standard Model states. Indicative explorations
of annihilation signatures may assume particular final states as illustrations,
but in the CMSSM these branching fractions are fixed at each point in parameter
space. We find that the dominant annihilation channels are $\tau^+ \tau^-$ in
the coannihilation region and $W^+ W^-$ in the focus-point
region for $\tan \beta = 10$, and ${\bar b}b$ in both regions for $\tan \beta = 55$.
The $\gamma \gamma$ final state, which would be particularly promising for
detection via a $\gamma$ line with $E_\gamma = m_\chi$, unfortunately has a
very small branching fraction in all cases. The next step is to model the hadronic, leptonic and $\gamma$
components of the final states, for which we use {\tt PYTHIA}, as
introduced at the end of  Section~2.

In Section~3, we turn to the treatment of the astrophysical background
and astrophysical aspects of indirect detection. In particular,
we discuss possible dark matter profiles towards the center of the galaxy,
including Navarro-Frenk-White (NFW)~\cite{NFW} and Einasto models~\cite{einasto,Graham},
as well as a simple isothermal model~\cite{iso}.
We also compare our predicted signal with a model background and the
current data and possible sensitivity of the Fermi-LAT detector~\cite{fermi} in Section~3,
illustrating our discussion with some specific benchmark CMSSM scenarios.
Then, in Section~4, we present estimates of the perspectives Fermi-LAT may have for
distinguishing a possible annihilation $\gamma$-ray
signature from the background along the WMAP strips shown in Fig.~\ref{fig:planes}. 
As might be expected on the basis of the
magnitudes of the annihilation cross sections in the different regions
of CMSSM parameter space, for $\tan \beta = 10$ there are better prospects for 
disentangling a signal from the background in the focus-point region,
whereas for $\tan \beta = 55$ there are prospects in both the funnel and
focus-point regions. However, putting pressure on the CMSSM scenarios
studied here with future years of Fermi-LAT data would require 
considerable improvements in the modelling of the background
and/or reduction in the systematic errors. Section~5 summarizes our conclusions
about searches for $\gamma$ rays from the galactic centre,
and makes some remarks about other search strategies for CMSSM
dark matter annihilations.

\section{CMSSM Dark Matter Annihilation Processes}

The relic density of neutralinos is determined by
annihilations when they are slightly non-relativistic.
Typically, annihilation freeze-out occurs when $x \equiv T/m_\chi \sim v_{\rm rel}^2/6  \sim 1/23$, 
where $v_{\rm rel}$ is the relative velocity of annihilating particles 
and $T$ is the temperature of the Universe. For small $x<1$, the annihilation cross section
can be expanded in a series as $\langle \sigma v_{\rm rel} \rangle = a + b x + \dots$ \cite{EHNOS,swo}. Because 
neutralinos are Majorana particles, we generally have $a \propto m_f^2/m_{\rm susy}^4$,
where $m_f$ is the mass of a final-state fermion, and $b \propto m_\chi^2/m_{\rm susy}^4$.
Hence, often $a \ll b$ and the relic density is largely determined by $b$, with higher-order
coefficients in the expansion relatively unimportant.
Any signal of dark matter annihilations will also be scaled by the
annihilation cross section. 
However, since dark matter particles  annihilating in the galactic halo today
are very non-relativistic with $v_{\rm rel} \ll 1$, the annihilations are essentially pure
S-wave, and $\langle \sigma v_{\rm rel} \rangle \approx a$.
This is a valid argument even close to  the inner region  of the  galactic halo, where the 
 N-body   simulations indicate that  the DM halo particles
are quite non-relativistic, with $v_{\rm rel} \sim \mathcal{O}(10^{-4}-10^{-3})$~\cite{NFW}.

We note in passing that in our calculation the Sommerfeld enhancement effect~\cite{hisano} is  not  important. 
One condition for having such a sizeable enhancement is a high degree of mass degeneracy 
between the LSP and another sparticle, such as the chargino or stau.
However, along the focus-point WMAP strip it is not possible to achieve the required 
amount of degeneracy between the LSP and lightest chargino and,
on the other hand,  along the stau coannihilation  strip the effect of the Sommerfeld 
enhancement is known not to be important~\cite{Hryczuk:2011tq}.

\begin{figure}[t]
\begin{center}
\epsfig{file=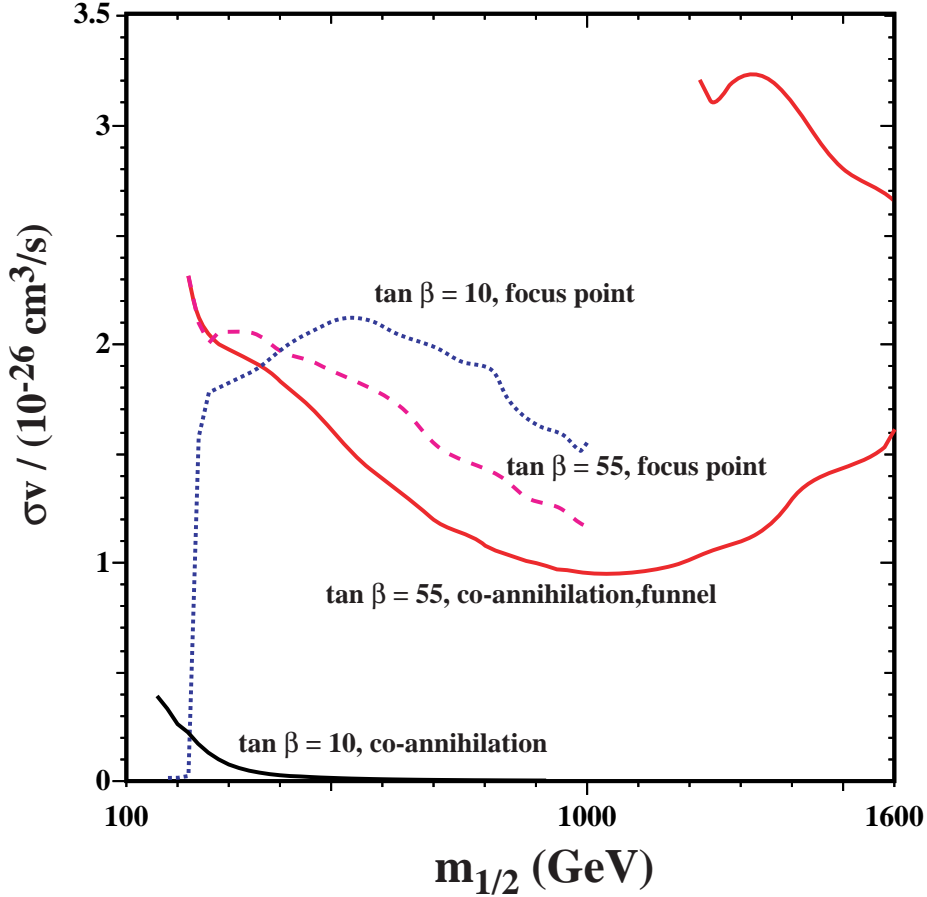,width=0.75\textwidth}
\caption{
\it The $\chi - \chi$ annihilation cross section along the WMAP strips in the
coannihilation, focus-point and funnel regions for $\tan \beta = 10, 55$,
$A_0 = 0$ and $\mu > 0$, as functions of $m_{1/2}$. We see that the annihilation
cross section along the $\tan \beta = 10$ coannihilation strip is much smaller
than along the other strips, and decreases rapidly as $m_{1/2}$ increases.}
\label{fig:strips}
\end{center}
\end{figure}

With this in mind, we display in Fig.~\ref{fig:strips}
the S-wave $\chi - \chi$ annihilation cross section along the WMAP-compatible
strips shown in Fig.~\ref{fig:planes}, as a function of $m_{1/2}$ in
each case. The coannihilation/funnel strips are represented by solid lines,
black for $\tan \beta = 10$ and red for $\tan \beta = 55$, and the
focus-point strips are represented by a blue dotted line for
$\tan \beta = 10$ and a mauve dashed line for $\tan \beta = 55$.
In each case, we have fixed $A_0 = 0$ and for each value of $m_{1/2}$
we have adjusted the value of $m_0$
to obtain the WMAP value of the relic density, $\Omega_\chi h^2 = 0.1109 \pm 0.0056$ \cite{WMAP}.

We notice immediately that the total $\chi -\chi$ annihilation
cross section is much smaller along the coannihilation strip
for $\tan \beta = 10$ than in the other cases, falling rapidly as $m_{1/2}$
increases. This reflects the importance of $\chi - {\tilde \ell}$
coannihilations in reducing the cosmological relic density into
the WMAP-compatible range, a role that becomes increasingly
important at larger $m_{1/2}$ where the $\chi$ and ${\tilde \tau_1}$
are increasingly degenerate. We recall that, as seen in Fig.~\ref{fig:planes},
the LHC upper limits on sparticle production enforce $m_{1/2} > 350$~GeV
along this strip, where $\sigma v < 10^{-27}$~cm$^3$/s, whereas $g_\mu - 2$
favours the portions of the coannihilation strips with $m_{1/2} < 400$~GeV
for $\tan \beta = 10$ and $m_{1/2} < 800$~GeV for $\tan \beta = 55$. 

On the other hand, the total S-wave
annihilation cross section is much larger along the other WMAP strips
shown in Fig.~\ref{fig:strips}, with typically $\sigma v \sim (1 - 2) \times 10^{-26}$~cm$^3$/s~%
\footnote{There is an exception at small $m_{1/2}$ on the $\tan \beta = 10$ focus-point strip, where the
annihilation $\chi \chi \to W^+ W^-$ is kinematically inaccessible.}.
This difference immediately suggests that detecting dark matter
annihilations will be easier along the funnel for $\tan \beta = 55$ as well as 
the focus-point strip for $\tan \beta = 10$,
or for $\tan \beta = 55$. We note, however, that these regions are disfavoured by fits to 
$g_\mu - 2$ and other low-energy precision observables~\cite{mc}.

In the case of the coannihilation/funnel strip for $\tan \beta = 55$, we note the
appearance of a second red line for $m_{1/2} > 1200$~GeV. This
reflects the appearance in Fig.~\ref{fig:strips} of a second branch of the WMAP strip
on the other side of the rapid-annihilation $H/A$ funnel seen in the right
panel of Fig.~\ref{fig:planes}.  
The annihilation cross section takes similar values along the focus-point
strips for both the $\tan \beta = 10$ and 55 cases (except at small $m_{1/2}$), and the LHC limits have
no impact along either of these strips. As seen in Fig.~\ref{fig:planes}, $g_\mu - 2$ favours only
the portion of the $\tan \beta = 55$ focus-point strip with $m_{1/2} < 200$~GeV, which is disfavoured
by other constraints~\footnote{We note that, for $\tan \beta = 55$, the LHC limits have less impact
on $m_{1/2}$ than the $b \to s \gamma$ constraint, which imposes
$m_{1/2} > 400$~GeV.},
and disfavours all the $\tan \beta = 10$ focus-point strip.

\begin{figure}[htb!]
\begin{center}
\epsfig{file=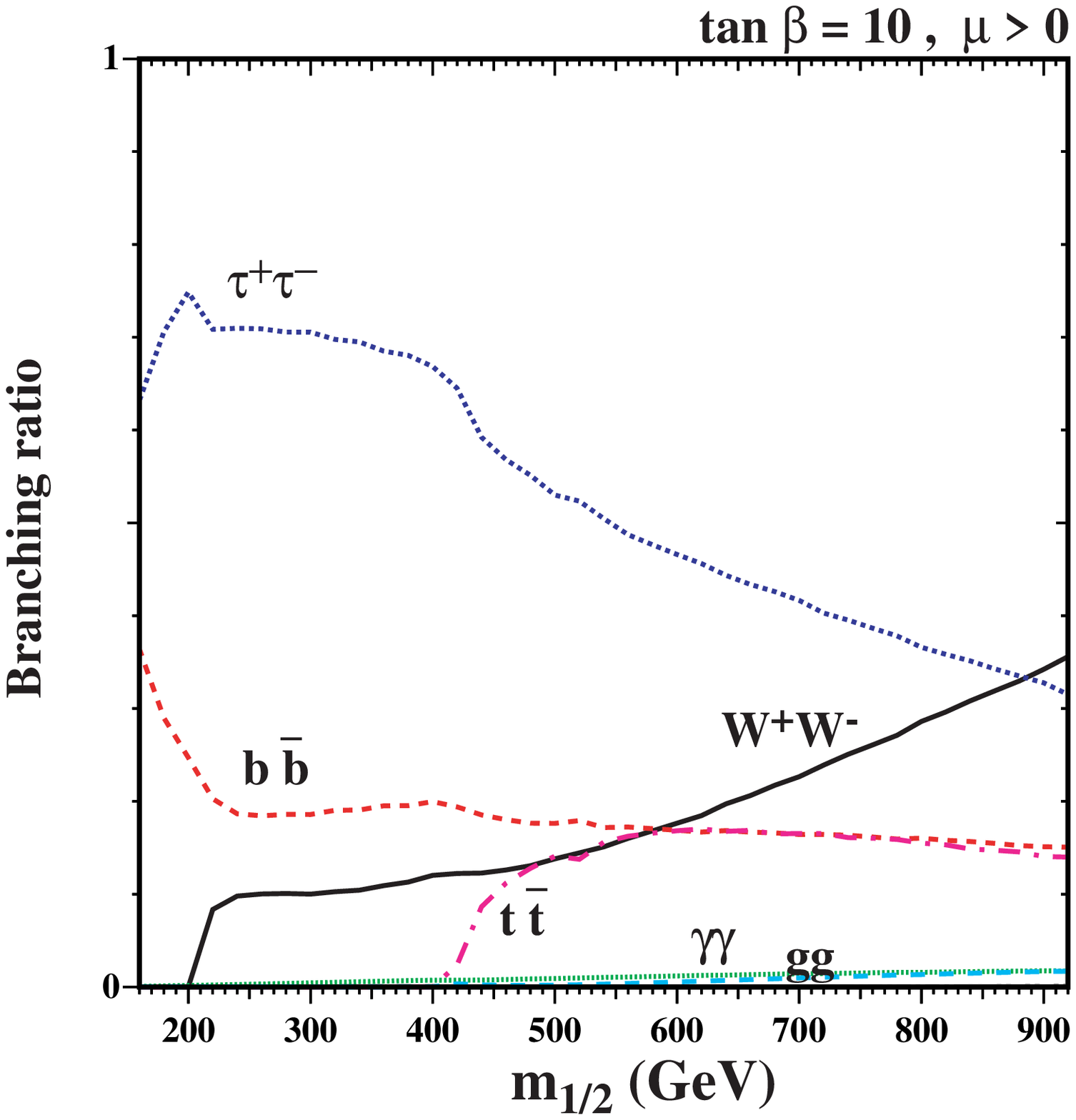,width=0.475\textwidth}
\epsfig{file=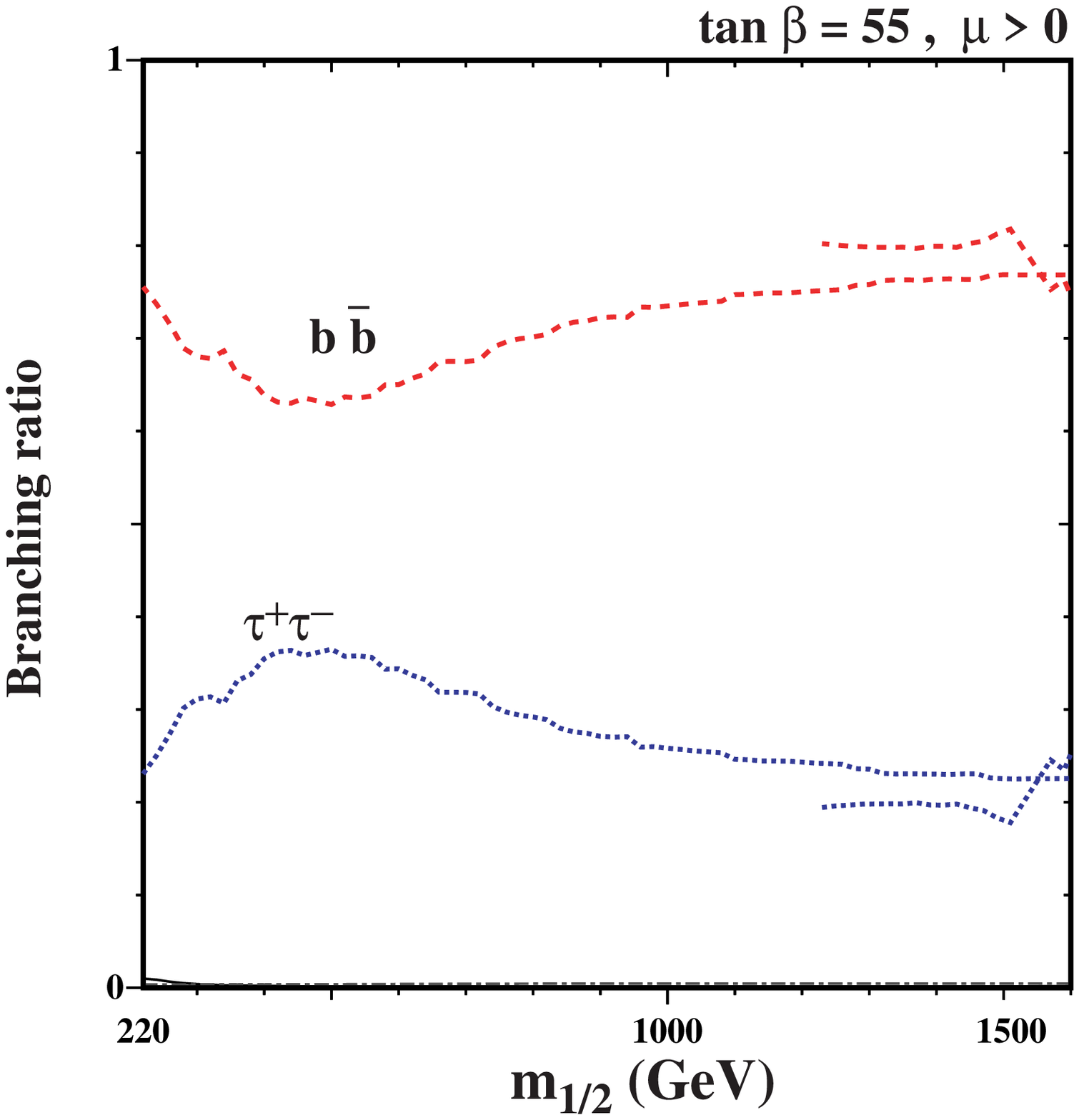,width=0.475\textwidth}\\
\epsfig{file=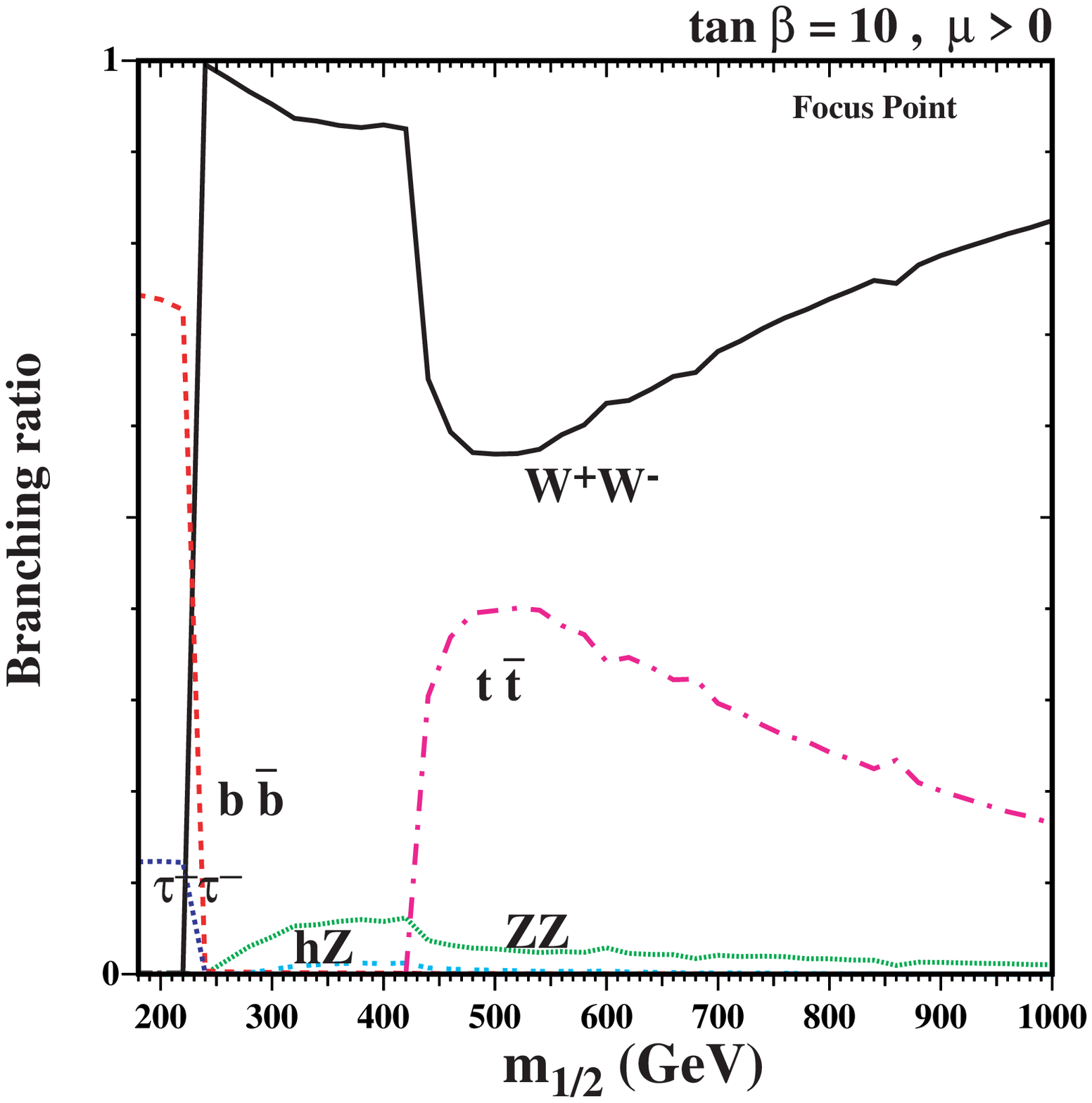,width=0.475\textwidth}
\epsfig{file=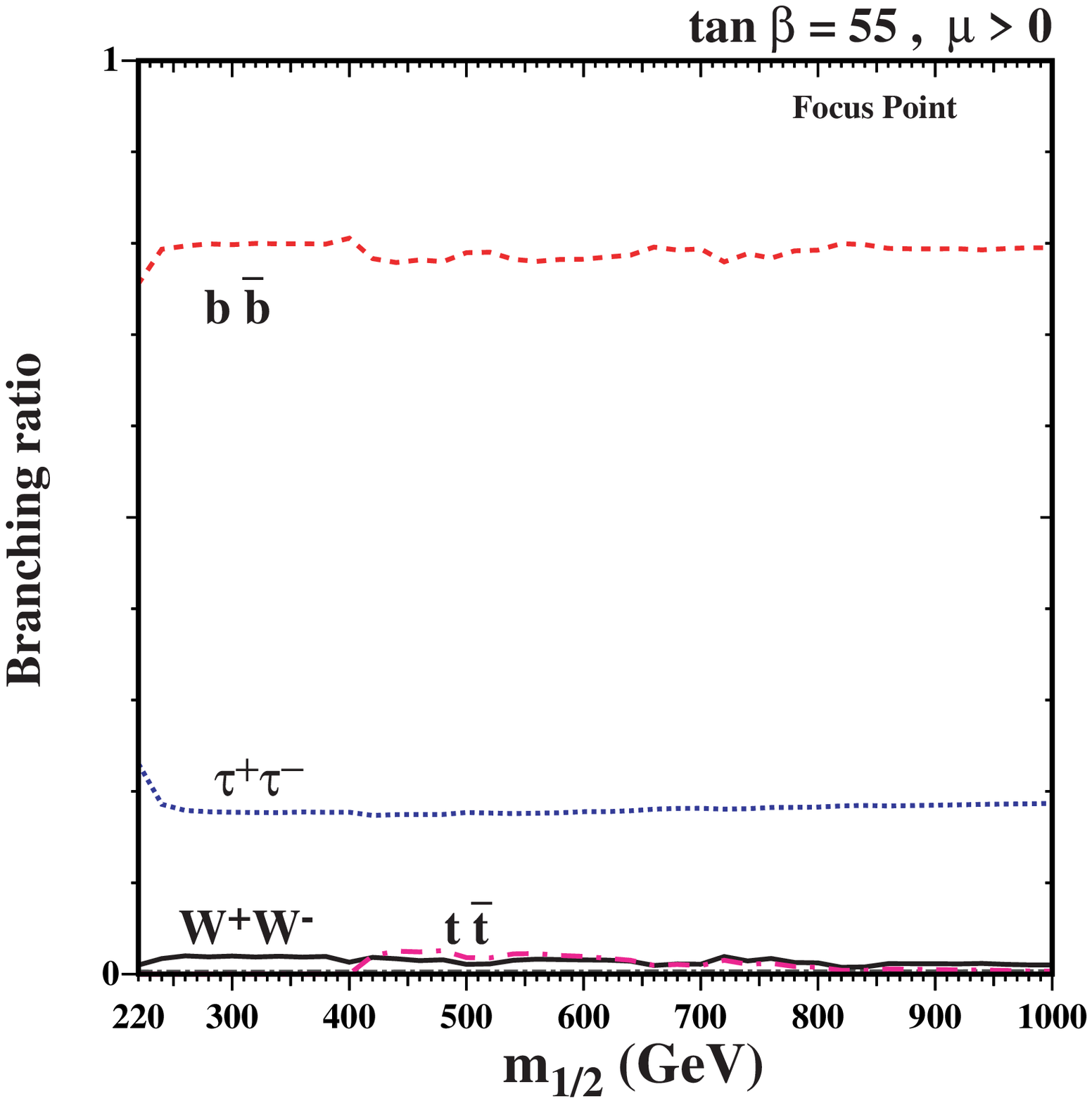,width=0.475\textwidth}
\caption{
\it The branching fractions for $\chi - \chi$ annihilations into pairs of
Standard Model particles along the WMAP strips for $\tan \beta = 10$
(left panels) and $\tan \beta = 55$ (right panels), in the coannihilation
and funnel regions (upper panels) and in the focus-point region
(lower panels).}
\label{fig:fractions}
\end{center}
\end{figure}

The detectability of $\chi - \chi$ annihilation depends also on the branching
fractions for annihilations into specific Standard Model final states and
the $\gamma$ spectra they produce. Fig.~\ref{fig:fractions} displays the
branching fractions for the most important final states, and we see that
they are quite different along the various WMAP strips studied. In the case
of the coannihilation strip for $\tan \beta = 10$ (upper left panel), we see
that $\tau^+ \tau^-$ final states dominate at low $m_{1/2}$, followed by
${\bar b} b$ final states, with $W^+ W^-$ and ${\bar t} t$ final states
gaining in importance at larger $m_{1/2}$, where the total annihilation cross
section is, however, much reduced as seen in Fig.~\ref{fig:strips}.

On the other hand, the roles of the ${\bar b} b$ and $\tau^+ \tau^-$
final states are largely reversed along the coannihilation/funnel strip
for $\tan \beta = 55$ (upper right panel), including the second branch when
$m_{1/2} \sim 1500$~GeV.
In the case of the focus-point strip for $\tan \beta = 10$ (lower left panel),
we see that $W^+ W^-$ final states dominate, except for ${\bar b} b$
final states at small $m_{1/2} \sim 200$~GeV and the appearance of
${\bar t} t$ final states when $m_{1/2} > 400$~GeV.
Finally, in the case of the focus-point strip for $\tan \beta = 55$ (lower right panel),
we see that ${\bar b} b$ final states dominate over $\tau^+ \tau^-$ final states
everywhere.

The prompt component of the photon flux, i.e., the photons that are produced directly 
by neutralino pair annihilation in the halo, consists of two components: the monochromatic and the continuum.
The monochromatic component due to the $\gamma \gamma$ branching fraction is very small in
all cases. For $\tan \beta = 10$, the branching fraction is less then $10^{-3}$ for 
$m_\chi \la 400$ GeV. It rises slowly to roughly 0.015 as the endpoint of the 
coannihilation strip is reached, but here the total annihilation rate is very small, as seen
in Fig.~\ref{fig:strips}. In the focus-point region for $\tan \beta = 10$,
the branching ratio is always negligible and remains below $10^{-7}$.
In the case of the coannihilation/funnel strip for $\tan \beta = 55$ it remains below a few $\times 10^{-5}$,
and is even lower along the focus-point strip for $\tan \beta = 55$.
Further details, especially for the monochromatic part from the processes 
$\chi \chi \to 2 \gamma / Z \gamma$, will be given in the next section.

The continuum component is due to photons  produced by the multibody neutralino pair annihilation channels,
and is usually  modelled using event generators such as {\tt PYTHIA} ~\cite{pythia} 
or {\tt HERWIG} \cite{herwig}.  Here we use {\tt PYTHIA} to calculate the continuum component of the 
differential photon flux $dN^i_{cont} /dE_\gamma$.
The decay channels we take into account are: $q \bar{q}$ for all quark flavors, and the $\tau^+ \tau^-$, 
 $W^+W^-$, $ZZ$, $g g$, $ Z h$ and $Z H$ final states~\footnote{In the regions of the 
 CMSSM parameter space 
 we study here, the $ZH$ final state, where $H$ is the heavy  CP-even Higgs boson, yields a very small contribution.}. 
 In particular, we follow~\cite{Cembranos:2010dm} and references therein,  parametrizing  the fits of photon  spectra 
 produced  by {\tt PYTHIA}, in order to obtain the differential flux as  an analytic function
 of the parameter $x=E_\gamma/m_\chi$, for each LSP mass $m_\chi$. 
 The functions $dN^i_{cont}/dx$ for
 each final state are related to the energy spectrum used below $dN^i_{cont}/dE_\gamma = (1/m_\chi)dN^i_{cont}/dx$.
Our results for the continuum component agree very well with those of~\cite{Cirelli:2010xx,Cembranos:2010dm}.
In the modeled fluxes we have included effects that are related to final state radiation~\cite{Bringmann:2007nk}.
Recently, it has been shown that  final state strong and EW corrections from the radiative 
emission of  gluons, $W$ or $Z$~ \cite{Ciafaloni:2010ti} can enhance  otherwise P-wave suppressed  channels,
like  the neutralino pair annihilation  to electrons and neutrinos.
In principle,  corrections to  the neutralino annihilation  to electrons can affect the  photon 
flux, due to  the Inverse Compton Scattering  (ICS) effect, but as  we will 
see in the following  section,  this  part of the signal is subdominant for energetic photons.

\section{Gamma fluxes and the Fermi-LAT data analysis}
\label{sect:fluxes}

In order to estimate the sensitivity of a detector like Fermi-LAT~\cite{fermi} to 
new sources of photon  fluxes arriving from  the galactic center (GC), one must calculate both the signal and the background
contributions to the photon flux, which we do in the following subsections. 

The GC is probably one of the most  complicated regions of the Milky Way,   from both the observational and the 
theoretical points of view. Observationally, the emitted electromagnetic spectrum exhibits 
various interesting features,  like the recently-observed
giant $\gamma$-ray bubbles, the so-called ``Fermi bubbles"~\cite{Su:2010qj}, that 
can be related  to accretion activity. 
From the theoretical point of view, the  supposed existence of a massive black hole in 
the GC might change drastically the size of the   $\gamma$-ray flux~\cite{smbh}.
Moreover, other phenomena such as turbulent
galactic winds, large magnetic fields etc,  can affect 
 the theoretical predictions~\cite{Crocker:2010qn}. 
In this work we adopt a conservative approach. In particular, we do not consider the possible effects 
 due to   the presence  of a massive black hole, or any other 
source that might enhance the $\gamma$-ray flux arising from annihilations
of neutralinos in the dark matter halo.

\subsection{The supersymmetric signal}

In the context of supersymmetric models where the neutralino plays the role of the dark matter particle,
there are two main components in the signal:
\begin{itemize}
 \item Prompt  photons produced either directly through one-loop processes such as
$2\chi \to 2 \gamma/Z\gamma$ (monochromatic components) or indirectly as final state radiation or through the hadronization, 
fragmentation and decays of the tree-level neutralino pair annihilation products discussed
in the previous section (the continuum component);
\item  Photons that originate  from energetic electrons and
positrons produced in neutralino pair annihilation via the  Inverse Compton Scattering  (ICS) process 
on background photons.
\end{itemize}
There is also a  synchrotron component produced by the energetic electrons and positrons
accelerated in  the galactic magnetic field, but 
it contributes at energies much  smaller than the prompt and ICS components~\cite{Bertone:2008xr},
below the effective Fermi-LAT threshold. Thus, we may write
\beq
{\frac{d \Phi_\gamma (E_\gamma)}{ d E_\gamma}}\Big\vert_{signal} =
                { \frac{d \Phi_\gamma (E_\gamma)}{d E_\gamma }}\Big\vert_{prompt} + {\frac{d \Phi_\gamma (E_\gamma) }{d E_\gamma }}\Big\vert_{ICS}\,,
                \label{eq:signal}
\eeq
and the total flux $\Phi_\gamma(E_{th})$ as a function of the  threshold energy $E_{th}$ is
 \beq
\Phi_\gamma(E_{th})=\int_{E_{th}}^{m_\chi} dE_\gamma  { \frac{d \Phi_\gamma  (E_\gamma)}{d E_\gamma }} \Big\vert_{signal} \, .
\label{eq:accuflux}
\eeq
In the following, we discuss in detail our calculation of the prompt and  the ICS components in the
supersymmetric $\gamma$ signal.

\subsubsection{The prompt component}

The prompt component may be written as
\beq
  { \frac{d \Phi_\gamma (E_\gamma) }{d E_\gamma }}\Big\vert_{prompt} =
\frac{1}{4\pi}  \frac{1}{2 \,m_\chi^2} \sum_X \frac{dN^X_\gamma}{dE_\gamma} {\langle \sigma \, v \rangle}_X
       \int_{\Delta \Omega} d \Omega \int_{\rm {los}} \rho^2(r(s,\psi)) ds ,
       \label{eq:prom}
\eeq
where we sum over the channels, $X$, of neutralino pair annihilation, both monochromatic  and continuum. 
The $ {dN^X_\gamma}/{dE_\gamma}$ are the photon energy spectra produced per 
annihilation event, for each channel with partial annihilation cross section $ {\langle \sigma \, v \rangle}_X$,
which we may separate as
\beq
\sum_X \frac{dN^X_\gamma}{dE_\gamma} {\langle \sigma \, v \rangle}_X=
2 \, v \sigma_{\gamma \gamma} \, \delta(E_\gamma-m_\chi) 
+  v \sigma_{Z \gamma} \, \delta \left(E_\gamma-m_\chi (1-\frac{M^2_Z}{4 m_\chi^2}) \right) +
\sum_i \frac{dN^i_{cont}}{dE_\gamma} {\langle \sigma \, v \rangle}_i \,.
\eeq
The first and the second terms correspond to the monochromatic channels $\chi \chi \to 2 \gamma / Z \gamma$, respectively,
for which we use the one-loop annihilation cross sections $ v \sigma_{\gamma \gamma}$ and  $v \sigma_{Z \gamma}$ 
computed in~\cite{Bergstrom:1997fh,Ullio:1997ke}.
For the continuum component   we calculate the $dN^i_{cont}/dE_\gamma$ 
using the partial cross sections $ {\langle \sigma \, v \rangle}_i $
for each of the two-body neutralino pair annihilation 
channels $i= \{q \bar{q}\, ,\tau^+ \tau^- \, , W^+W^- \, ,ZZ \, ,g g \, , Z h \}$, multiplied by
the corresponding continuum spectra modelled using {\tt PYTHIA}.
 
We follow the standard procedure~\cite{Bergstrom:1997fj} for the halo factor in (\ref{eq:prom}),
defining a dimensionless average halo factor as an integral along light-of-sight  (los) directions
within a solid angle $\Delta \Omega$: 
\beq
\bar{J}(\Delta \Omega)  \,  \Delta \Omega \equiv \frac{1}{R_\odot \rho_\odot^2}  \int_{\Delta \Omega} d \Omega \int_{\rm {los}} \rho^2(r(s,\psi)) ds \, .
\label{eq:Jav}
\eeq
where $R_\odot$ is the solar distance from the GC and $ \rho_\odot$ is local dark matter density.
With these definitions, the prompt part in (\ref{eq:prom}) takes the form 
\beq
 { \frac{d \Phi_\gamma (E_\gamma) }{d E_\gamma }}\Big\vert_{prompt} =
\frac{R_\odot }{2\pi}  \left( \frac{ \rho_\odot}{2 \,m_\chi} \right)^2  \, 
(\bar{J} (\Delta \Omega)  \Delta \Omega) \, \sum_X \frac{dN^X_\gamma}{dE_\gamma} {\langle \sigma \, v \rangle}_X  \, .
   \label{eq:prompt}
\eeq
We use the values  $R_\odot=8.5$ kpc and  $ \rho_\odot=0.3\, \rm{GeV/cm^3}$.
The  integration along the  light-of-sight direction can be performed using
\beq
 r^2(s,\psi)=R_\odot^2-2sR_\odot\cos\psi+s^2 ,
 \eeq
where $\psi$ is the angle with respect to the GC direction.
Defining the function  
\beq
J(\psi) \equiv \frac{1}{R_\odot \rho_\odot^2}  \int_{0}^{s_{max}(\psi)} \rho^2( \sqrt{R_\odot^2-2sR_\odot\cos\psi+s^2 } )  ds\, ,
\label{eq:Javer}
\eeq
one can calculate the halo average in a cone within angle $\phi$, that defines the size of the 
observation window  around the GC, as
\beq
\bar{J}(\Delta \Omega)  \,  \Delta \Omega = 2 \pi \int_{\cos\phi}^1 J(\psi) d(\cos \psi) \, .
\eeq
This is the quantity introduced in (\ref{eq:Jav}),
where we have used $\Delta \Omega=2 \pi (1 -\cos\phi)$. The upper limit
$s_{max}$ in (\ref{eq:Javer}) is calculated  in terms of the
Milky Way halo size $R_{\rm{MW}}$
\beq
s_{max}(\psi)=\sqrt{R^2_{\rm {MW}}  - ( R_\odot \sin\psi )^2 } + R_\odot \cos\psi 
\eeq 
although, as noted in ~\cite{Yuksel:2007ac}, the contribution of the
integration beyond the scale radius $\sim 20 - 30$ kpc is negligible.

The behavior of the galactic DM halo, especially in the  inner region, is a very controversial topic.
Although some  N-body simulations indicate that  the  halo 
presents a highly cusped behavior towards its center~\cite{NFW,Moore:1999gc},
other simulations ~\cite{iso,Burkert:1995yz} as well as  atomic hydrogen H{\scriptsize I} observations on
 dwarf spiral galaxies~\cite{spiral} indicate   shallower profiles.
Recently, the  Via Lactea 2  simulations~\cite{Diemand:2008in} seem to  support a cuspy  profile like  NFW,
 whereas  the Aquarius DM simulation project~\cite{Springel:2008cc}  favors a 
 different parameterization~\cite{Graham,Navarro:2008kc}, that 
does not have such a cuspy character in the inner region, such as the Einasto profile.
This unsettled issue of the cuspiness 
has significant impact on the theoretical predictions
 for  the $\gamma$-ray flux emitted from  the GC,  and  
suggests a substantial theoretical uncertainty in calculating this.
To address this uncertainty,  
we study three  halo profiles that behave differently in the inner region of the
 Milky Way:  NFW~\cite{NFW}, Einasto~\cite{einasto,Graham}  and a simple isothermal profile~\cite{iso}. 
The NFW  behaves like $r^{-1}$ at small distances, while 
 the Einasto and isothermal profiles are both non-singular towards the galactic center.

These profiles are defined by the density functions:
\bea
\rho_{\rm{NFW}}(r)&=& \rho_s \frac{r_s}{r} \left( 1+\frac{r}{r_s}  \right)^{-2} \, , \nonumber  \\ 
\rho_{\rm{Ein}}(r)&=& \rho_s \exp \left[  -\frac{2}{\alpha} \left(   \left(  \frac{r}{r_s}  \right)^\alpha   -1  \right)        \right] \,  , \nonumber  \\
\rho_{\rm{iso}}(r)&=& \frac{\rho_s}{1+(r/r_s^\prime)^2}  \,  .
\eea
We choose the following values of the model parameters: $ r_s=20$ kpc, $r_s^\prime= 5$ kpc and $\alpha=0.17$.
The $\rho_s$ parameter, which has a different numerical value for  each profile,
 is  defined in such a way that $\rho_\odot = \rho(R_\odot)=0.3 \, \rm{GeV/cm^3}$.
Assuming these numerical values, values of the halo factor $\bar{J}(\Delta \Omega)  \,  \Delta \Omega$
 for three windows around the GC: 10, 7 and 5 degrees are tabulated in Table~\ref{table:halo}.
The values in this Table are useful for understanding the relative sizes of the halo factor for
the three halo profiles we study and the various windows. 
In the case of the the  prompt component, the halo factor is simply multiplicative.
 In the case of the the  ICS component,  the  halo factor  is also, to a good approximation, multiplicative~\cite{Cirelli:2010xx}, but 
 there are also other terms related to the photon background parameters and to
 the propagation of the lepton fluxes, etc.,  that are different 
 for the various halo profiles and windows. 

 \begin{table}[ht!] 
\centering 
 \begin{tabular}{crrr} 
 \hline  \hline     
 Model &  10 deg & 7 deg & 5 deg \\ 
 \hline
 NFW          & 10.51   & 7.90     & 5.95     \\ 
Einasto       & 19.68   & 15.21     & 11.56 \\
Isothermal  & 1.21     & 0.62        & 0.32   \\ [1ex] 
\hline
\end{tabular}
 \caption{\it Values of the halo factor $\bar{J}(\Delta \Omega)  \,  \Delta \Omega$ for various galactic dark matter halo profiles and
 values of the angle $\phi$ that defines the size of the observation  window   around the GC.}  
 \label{table:halo}
 \end{table}

\subsubsection{The Inverse Compton Scattering component}

Neutralino pair annihilation produces energetic electron and positron  ($e^\pm$) fluxes
mainly indirectly through the hadronization, fragmentation and decays of the primary annihilation products~%
\footnote{The $e^\pm$ fluxes produced directly through the channels $2 \chi \to e^+e^-$  are negligible,
because this process is P-wave suppressed and proportional to the small electron mass.}.
The indirectly produced $e^\pm$ scatter on the ambient photon background, and through 
the ICS effect produce energetic photons. In order to calculate the ICS part we follow 
the tools and methods described in~\cite{Cirelli:2010xx,Bernal:2010ip}.  

The ICS flux is given by the equation 
\bea
{\frac{d \Phi_\gamma (E_\gamma) }{d E_\gamma }}\Big\vert_{ICS} & = &  \frac{1}{E_\gamma^2} \, 
     \frac{R_\odot }{2\pi}  \left( \frac{ \rho_\odot}{2 \,m_\chi} \right)^2  \,  \nonumber  \\ 
 &\times&  \int_{\Delta \Omega} d \Omega
   \int_{m_e}^{m_\chi} d E_s  \sum_i \frac{dN^i_{e^\pm}(E_s)}{dE}\,  {\langle \sigma \, v \rangle}_i \, I_{IC}(E_\gamma,E_s, \psi) \, ,
\eea
where $E_s$ is the $e^\pm$ injection energy
and  the halo function for the inverse Compton radiative process $ I_{IC}(E_\gamma,E_s, \psi)$ is  defined
as
\bea
I_{IC}(E_\gamma,E_s, \psi)  &=& 2 E_\gamma \int_{los} 
                  \frac{ds}{R_\odot} \left(   \frac{\rho(r(s,\psi))}{ \rho_\odot} \right)^2 \nonumber  \\ 
&\times &\int_{m_e}^{E_s} d E \, \frac{\sum_a \mathcal{P}^a_{IC}(E_\gamma,E,r(s,\psi))  }{b(E,r(s,\psi))} I(E,E_s,r(s,\psi)) .
\eea
We calculate the primary electron/positron fluxes $dN^i_{e^\pm}/dE$ produced by neutralino pair annihilation in the halo
using {\tt PYTHIA}, and the index $i$ runs over the channels used for the photon prompt part. 
The functions $ \mathcal{P}^a_{IC} $, 
the energy loss coefficient function $ b(E,r)$
and the generalized halo function $I(E,E_s,r(s,\psi))$
 are given in~\cite{Cirelli:2010xx} and references therein.
 
The energy loss coefficient $b(E,r)$ depends on the profile 
of the magnetic field in the galactic plane, as described in~\cite{Strong:1998fr,Cirelli:2010xx}.
Recent estimates suggest that the  magnitude of the magnetic field towards the GC
can be larger by  a factor of ten or more than in the solar neighbourhood~\cite{Crocker:2010xc}. 
We have verified that a magnetic field that is stronger by a factor of 10 would result in a reduction of the ICS flux 
almost by a factor of two. This would affect the total $\gamma$-ray flux at low photon energy, where the ICS is dominant.
However, since the most important part of the CMSSM $\gamma$ signal is at higher energies, its 
effect on the global $\chi^2$ that we calculate in the next Section is small, of the order of 10\%. A similar remark applies
to the possibility of energy losses due to interactions with gas in the GC~\cite{Ferriere}.

The sum over the index $a$  includes three photon backgrounds: the cosmic microwave background (CMB), 
the starlight in the galactic plane and the infrared radiation due to the rescattering of the starlight by dust. 
To calculate the ICS part, we follow the semi-analytic method described in~\cite{Cirelli:2010xx}, see 
also~\cite{Delahaye:2007fr,Bernal:2010ip}, which yields results similar to numerical methods such as GALPROP~\cite{galprob}. 
In particular, for the $e^\pm$ propagation parameters there are three 
commonly used models, called MIN, MED and MAX~\cite{Delahaye:2007fr},
that correspond to the minimal, median and maximal primary positron fluxes compatible with data on the boron to carbon 
ratio,  B/C~\cite{Maurin:2001sj}.
 In this paper we choose the MED set of parameters, but we have checked that our findings  do not depend significantly
 on this choice. Specifically,  scanning through the various parameter sets, we find 
 that the value of the ICS flux can be changed by at most 15\%.  In the low-energy range, 
 this uncertainty affects  analogously the total $\gamma$-ray signal,
 but its effect in  the statistical $\chi^2$ analysis in Section~\ref{expsect} is unnoticeable.

\subsection{Examples within the CMSSM}

\begin{figure}[t] 
\epsfig{file=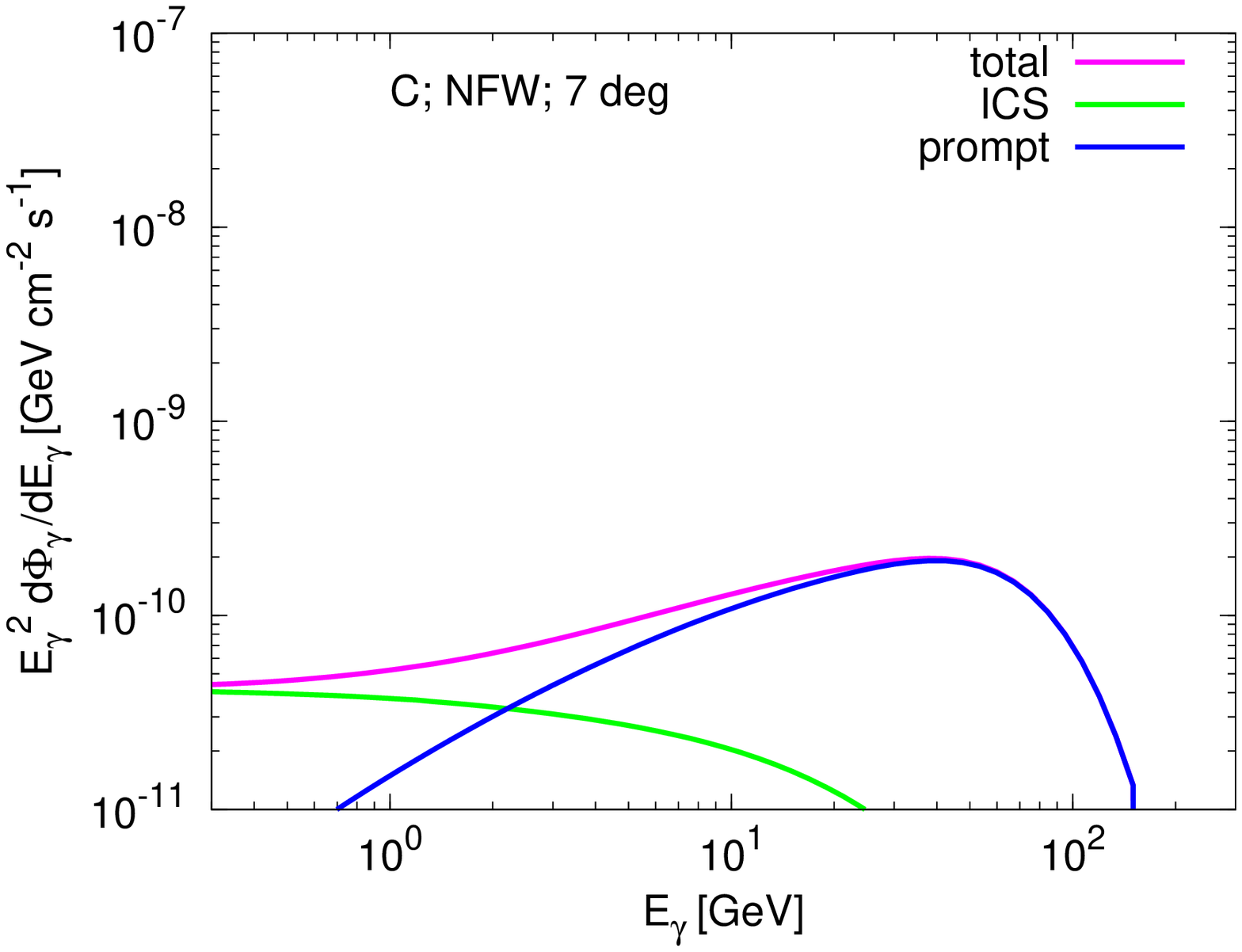,width=0.475\textwidth}
\epsfig{file=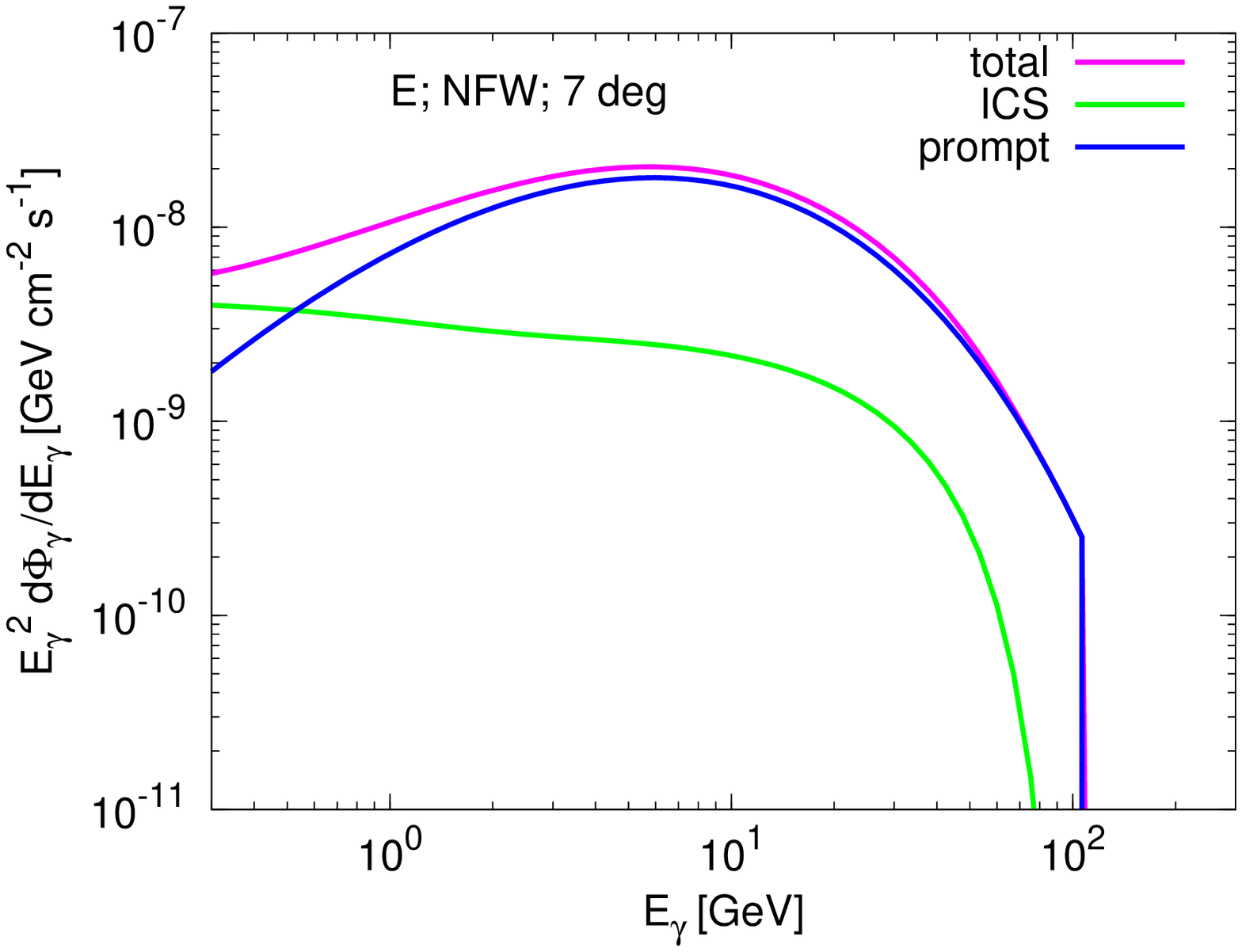,width=0.475\textwidth}\\
\epsfig{file=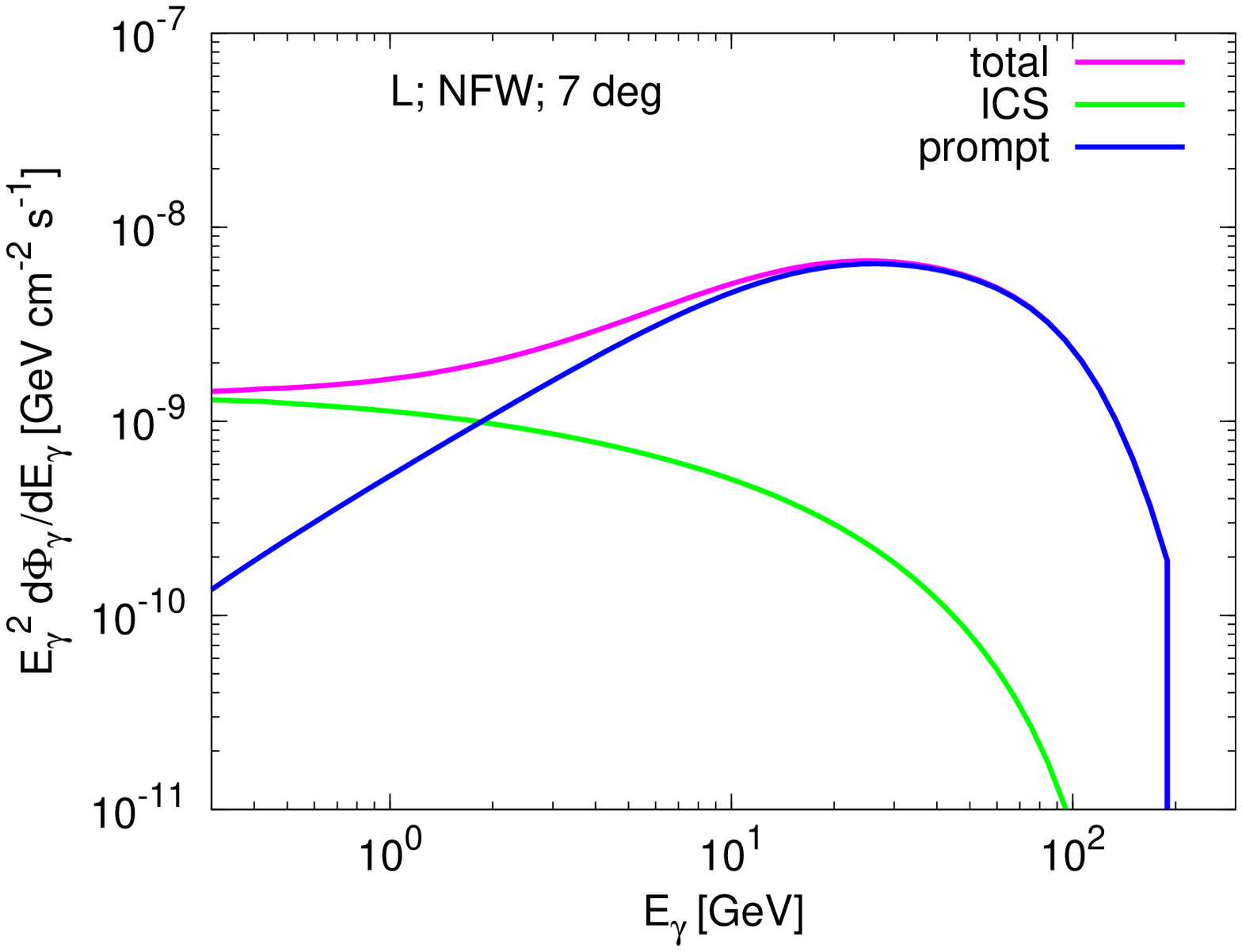,width=0.475\textwidth}
\epsfig{file=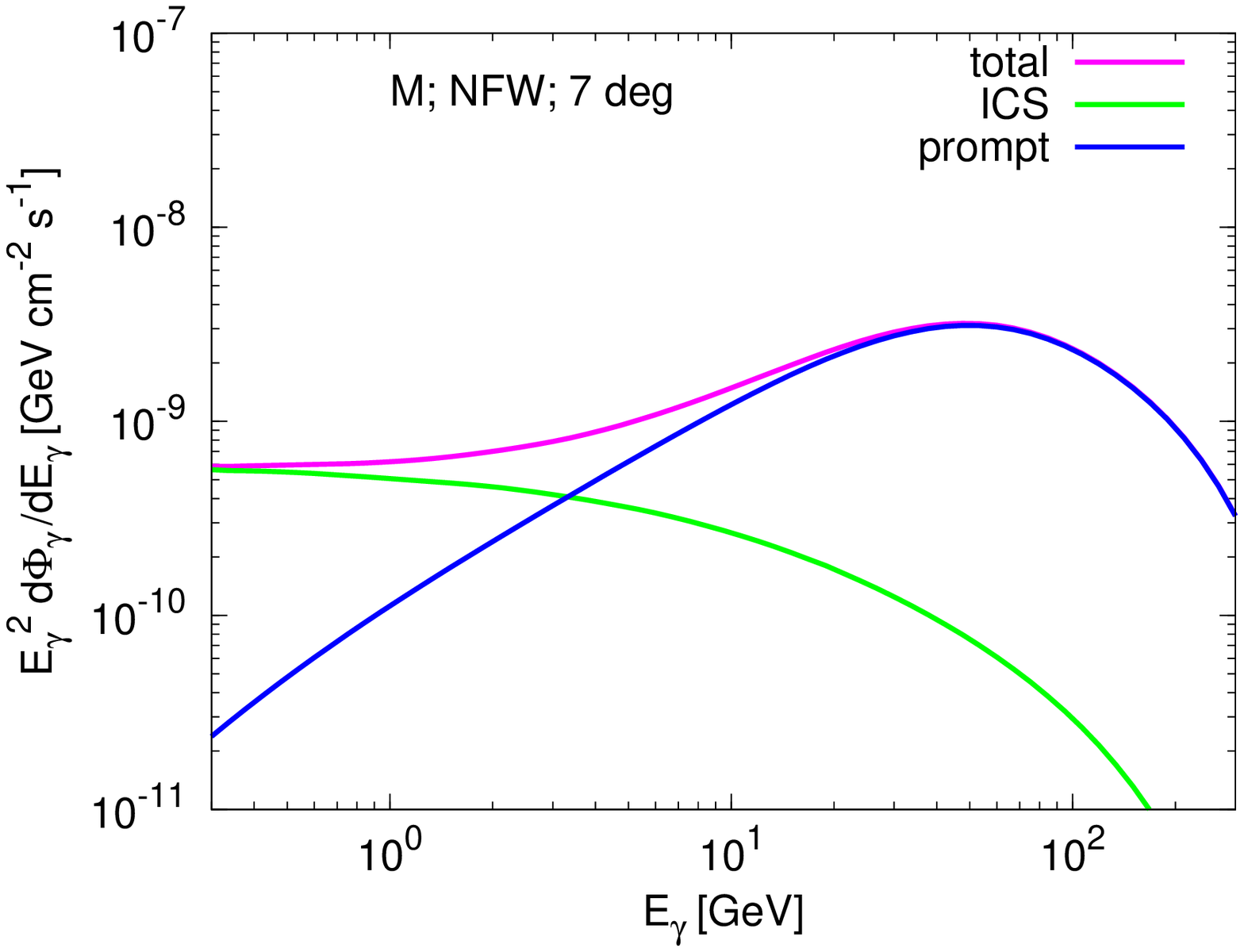,width=0.475\textwidth}
 \caption{\it The differential flux $E^2_\gamma \, d \Phi_\gamma/dE_\gamma$ as a function of the photon 
  energy $E_\gamma$  for a 7 deg window around the GC.  The blue curve is the 
  prompt part and the green curve is the ICS flux. We use here the NFW profile, and the four
  panels correspond  to the four benchmark CMSSM points C,E,L and M, which are described in the text.}
  \label{fig:fluxes}
\end{figure}

Using the above treatments of the prompt and the ICS parts,  
we can evaluate the total flux in (\ref{eq:signal}) and (\ref{eq:accuflux})
at any point of the parameter space of the CMSSM. 
In Fig.~\ref{fig:fluxes}, we display examples of 
the differential flux $E^2_\gamma \, d \Phi_\gamma/dE_\gamma$ as a function of the photon 
energy $E_\gamma$ in a 7-degree window around the GC.  The blue curves are the 
  prompt parts and the green curves correspond to the ICS contributions. 
  We use the NFW profile in each case, and the four panels in Fig.~\ref{fig:fluxes} correspond 
  to the four benchmark points C,E,L and M introduced in~\cite{bench} and updated in~\cite{eosgs}. 
  These points belong to different characteristic regions of the CMSSM model that 
  are  cosmologically acceptable.

Referring to the left panel of Fig~\ref{fig:planes} for $\tan \beta = 10$,
point C represents the low-mass region of the coannihilation strip for $\tan \beta = 10$ with 
$(m_{1/2},m_0)$ = (400,96) GeV,
and point E is in the focus-point region with $(m_{1/2},m_0)$ = (300,2003)~GeV 
(outside of the range shown in Fig~\ref{fig:planes}a).
Point L is in the coannihilation region at $(m_{1/2},m_0)$ = (450,312)~GeV and $\tan \beta = 50$,
and referring to the right panel of Fig~\ref{fig:planes} for large $\tan \beta$, point M is in the
rapid-annihilation funnel region at large $(m_{1/2},m_0$ = (1075,1045)~GeV with $\tan \beta = 55$.
The LSP masses at the points C,E,L and M  are $m_\chi=165, \, 117, \, 193 $ and $474 \gev $, respectively. 
Since point C is in the coannihilation region, it has a small neutralino annihilation cross section, 
as we saw in Fig.~\ref{fig:strips},
and hence the photon flux for point C  is weaker than for the other benchmark points. 
The focus-point region point E
has a stronger  flux, because it has a larger annihilation cross section and 
small $m_\chi$. 
The points L and M have comparable fluxes, but that of point M is relatively smaller,  because there  $m_\chi$ is larger~%
\footnote{It is useful to note here that both the prompt and ICS fluxes are, in general, 
inversely proportional to $m_\chi^3$.}.

  \begin{figure}[t] 
\epsfig{file=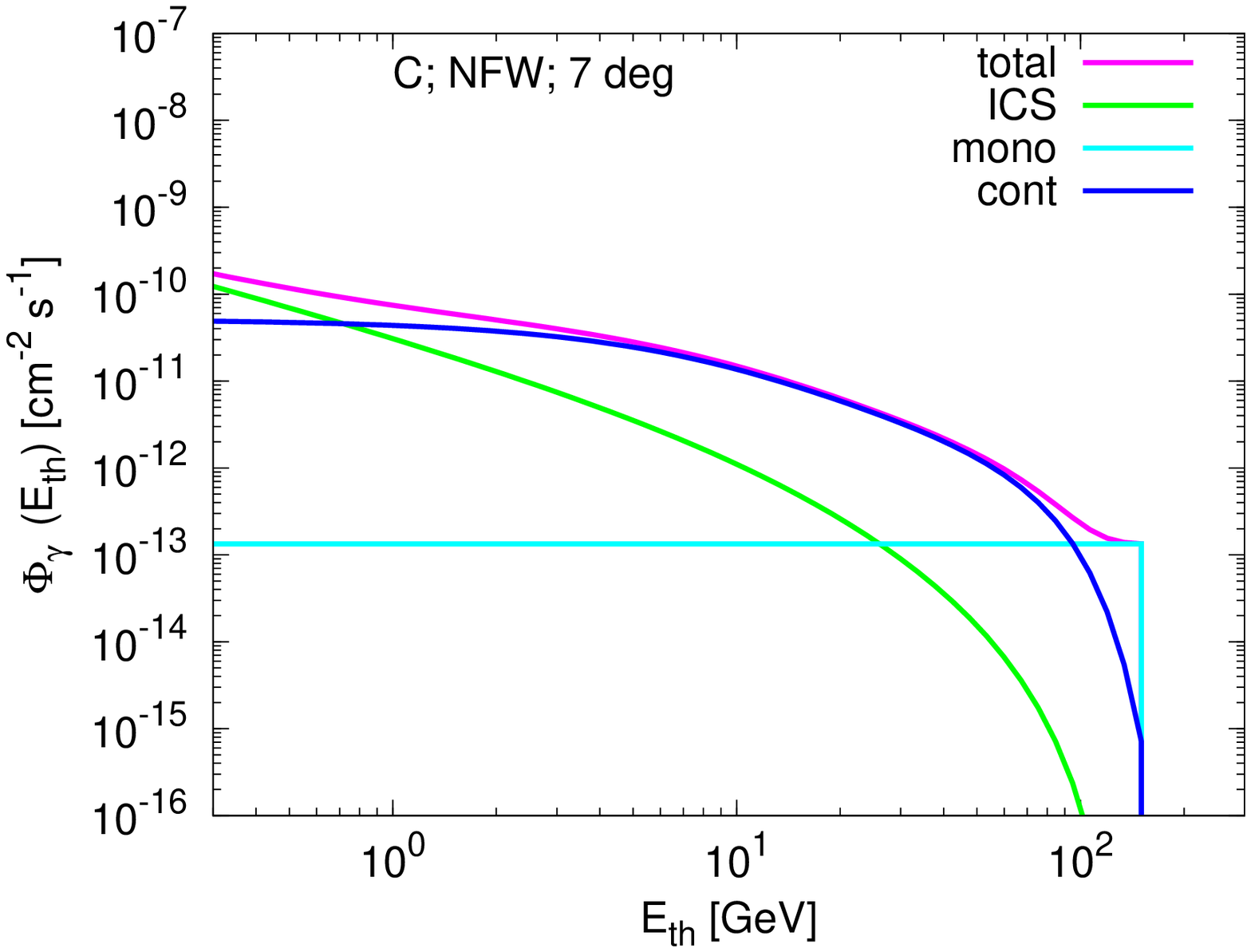,width=0.475\textwidth}
\epsfig{file=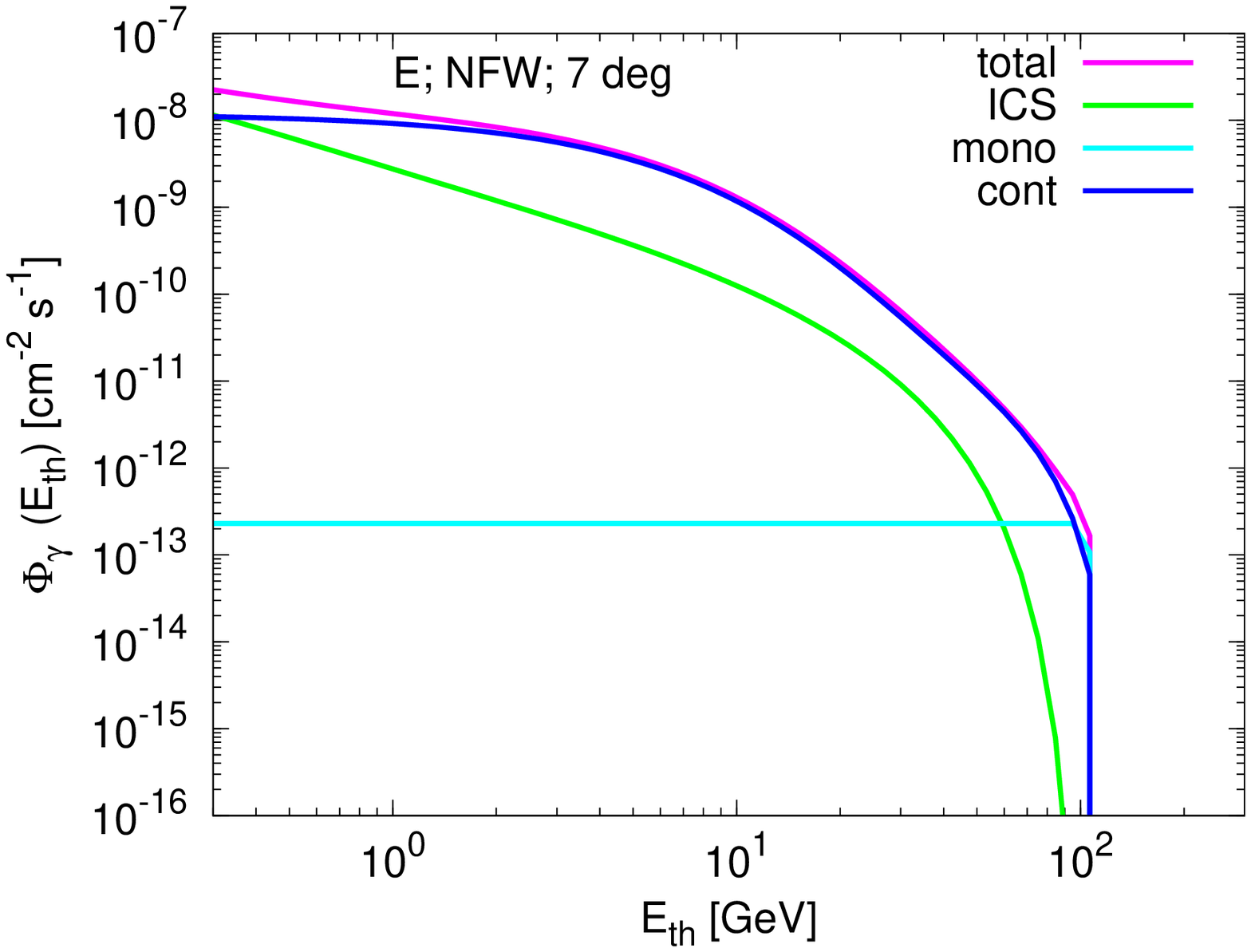,width=0.475\textwidth}\\
\epsfig{file=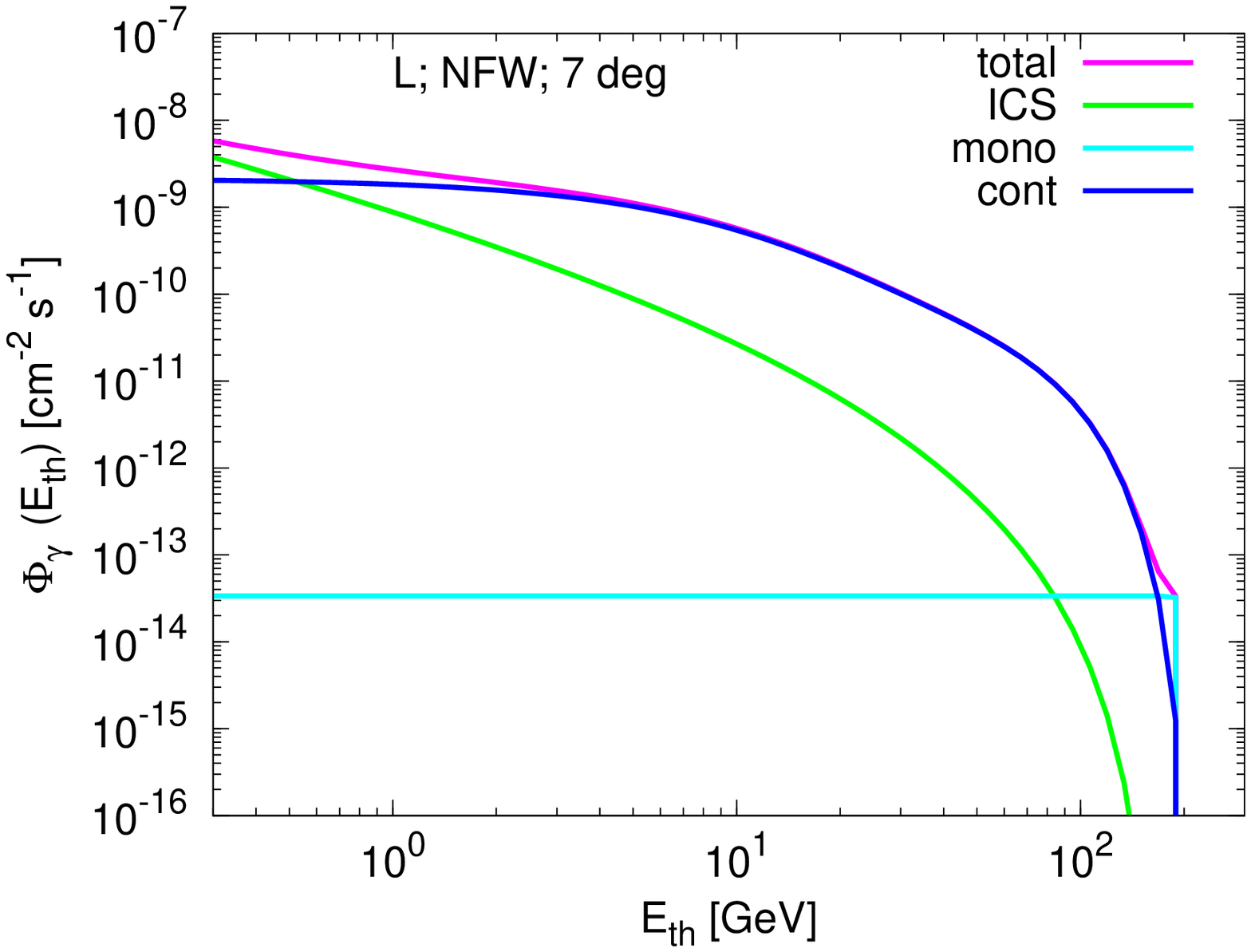,width=0.475\textwidth}
\epsfig{file=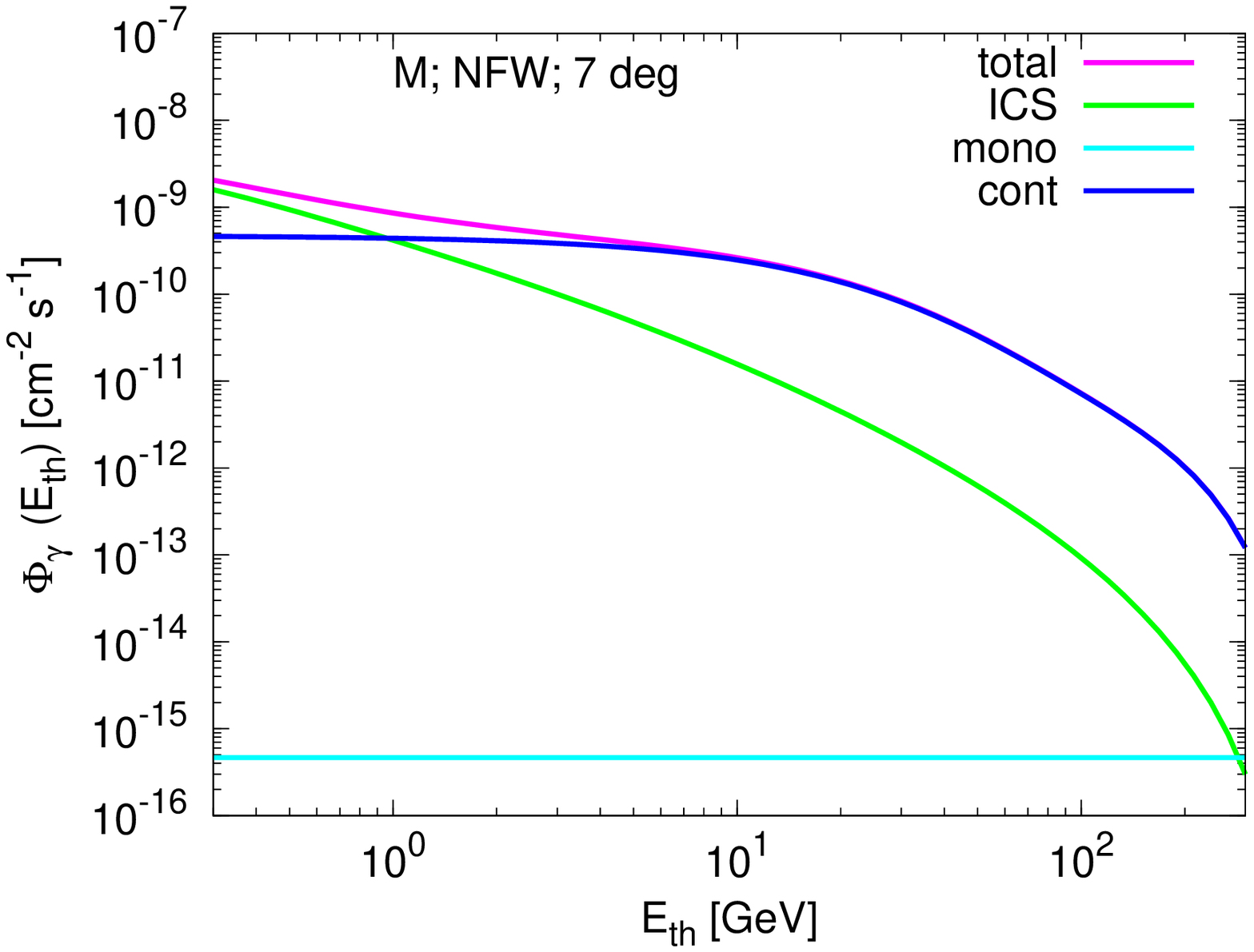,width=0.475\textwidth}

  \caption{\it The total flux $\Phi_\gamma(E_{th}) $ as a function of the threshold energy $E_{th}$ for a 7 deg window around the GC. 
  The green curves represent the ICS part, the cyan curves represent the monochromatic part, and the blue curves
  the continuum part. We use here the NFW profile, and the four
  panels correspond  to the four benchmark CMSSM points C,E,L and M, as in   Fig.~\ref{fig:fluxes}.}
  \label{fig:acc_flux}
\end{figure}

In Fig.~\ref{fig:acc_flux}, we plot the fluxes $\Phi_\gamma(E_{th})$  in a 7-degree window around the GC,
for the same benchmark  points as functions of the threshold energy $E_{th}$, as defined in (\ref{eq:accuflux}). 
Once again, the green curves represent the ICS parts, the cyan curves represent the monochromatic parts and the blue
curves  the continuum parts. We see in these figures that the monochromatic component yields only
a very small contribution in every case, as was discussed in the previous Section,
and is also inversely proportional to $m_\chi^2$.
Another remark concerns the relative magnitudes of the ICS and prompt components.
Since the ICS component results from background photons energized by interaction with 
energetic charged leptons, it dominates for  photon energies much smaller than $m_\chi$.
 In particular, for  LSP masses up to a few hundred  GeV as along the benchmark strips we study, 
the   ICS contribution becomes subdominant above the range $E_\gamma \gtrsim 1-3 \gev$, 
as can be seen in  Fig.~\ref{fig:fluxes}.

\subsection{The background and Fermi-LAT data}

The background has three main components: diffuse galactic emission (DGE), an isotropic extragalactic contribution 
and resolved point sources. 
DGE is produced essentially by the interaction of cosmic-ray nucleons and electrons 
with the interstellar medium. Nucleon interactions produce photons via $\pi^0$ decays, whereas 
electrons produce DGE via  bremsstrahlung and ICS.
For the determination of the DGE background component, a 
conventional model was developed~\cite{dge} using the 
hypothesis that the cosmic-ray spectra
in the Galaxy can be normalized to solar-system measurements.
This model did not reproduce well the EGRET data, especially for $E_\gamma \sim \mathcal{O}(\gev)$,
resulting in the apparent EGRET excess~\cite{Hunger:1997we}.
An updated version of the model that was able to describe the data was developed in~\cite{Strong:2004ry},
and the recent Fermi-LAT measurements on the diffuse background component~\cite{dgelat}
are well reproduced by the updated version of the conventional model. This is the diffuse model that 
we use in our analysis. As was pointed out in ~\cite{Strong:2004ry}, to estimate the errors in this model is a complicated task,
but based on the quality of its fit to the data Fermi-LAT, 
we can say that is about 20\% -- 25\% for photon energies up to 100 GeV.

 The isotropic diffuse emission or isotropic gamma-ray background (IGRB) 
is much fainter and, although 
the term extragalactic gamma-ray background is usually applied, its extragalactic origin is not universally accepted.    
Various astrophysical objects can contribute  to this emission, such as active galactic nuclei,  galaxy clusters,
  blazars,  star-forming galaxies, ultra-high-energy cosmic rays, etc.~\cite{igrb}.
The Fermi-LAT collaboration  measurement of the IGRB~\cite{igrblat} is  consistent with a power law,
somewhat  softer than  the previous EGRET analogous measurement~\cite{igrbegret}. In any case,
the isotropic component is subdominant in our analysis. 

On the other hand, resolved point sources (RPS) provide an important part of the photon background 
from the direction of the GC. We use the first 11-month Fermi-LAT catalogue~\cite{rpslat},
which contains 1451 point sources
modelled in the energy range 100 MeV to 100 GeV. The gamma flux of each point source is taken
to obey a simple power law.

In order to evaluate the possible constraints imposed on the CMSSM parameter space by the latest 
Fermi-LAT data~\cite{fermi}, we estimate the background using the Fermi-LAT Science Tools~\cite{ST}.
In our analysis, we use data collected by Fermi-LAT between Aug 4,  2008 and April 29, 2011,
making a selection based on the  recommendations of  the collaboration. We focus our analysis 
in the energy range 300 MeV to 300 GeV, dividing it in 25 bins spaced evenly on a logarithmic scale.   
We have studied various windows around the GC in the range 1--10 degrees, but we 
choose as the basic region-of-interest (ROI) of our analysis the 7-degree window centered  
at the position of the brightest source in the GC: 
$\mathrm{RA}=266.46^0$, $\mathrm{Dec}=-28.97^0$, as in~\cite{Vitale:2009hr}. Note however,
that our conclusions are not overly sensitive to this choice of window size.

\begin{figure}[t!]
\centering \epsfig{file=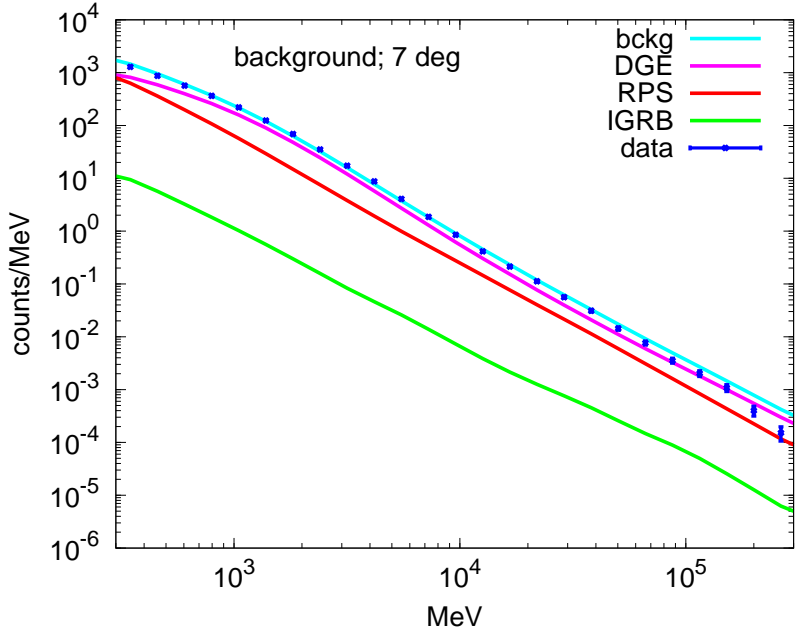,width=0.7\textwidth}
  \caption{\it
    The various background components in counts per MeV for a 7 deg window around the GC
    in the range 300 MeV to 300 GeV, estimated using the Fermi-LAT Science Tools~\protect\cite{ST}.
   We also display the Fermi-LAT data~\protect\cite{fermi} collected during the period described in the text. }
  \label{fig:latbckg}
\end{figure}

Fig.~\ref{fig:latbckg} displays the various background components obtained using the Science Tools~\cite{ST},
along with the data collected during the aforementioned period. The error bars attached to the data
represent the purely statistical errors. We have performed a binned likelihood analysis using the {\tt gtlike} tool, including 
in the background model file used in the fit  1) the galactic diffuse model,  2) the isotropic spectral template, 
and 3) the point sources, all as
provided by the collaboration. Concerning the point sources, we have included additional RPS from the vicinity of the 
ROI, that possibly  can contribute to the observed counts. 

\begin{figure}[t!]
\centering \epsfig{file=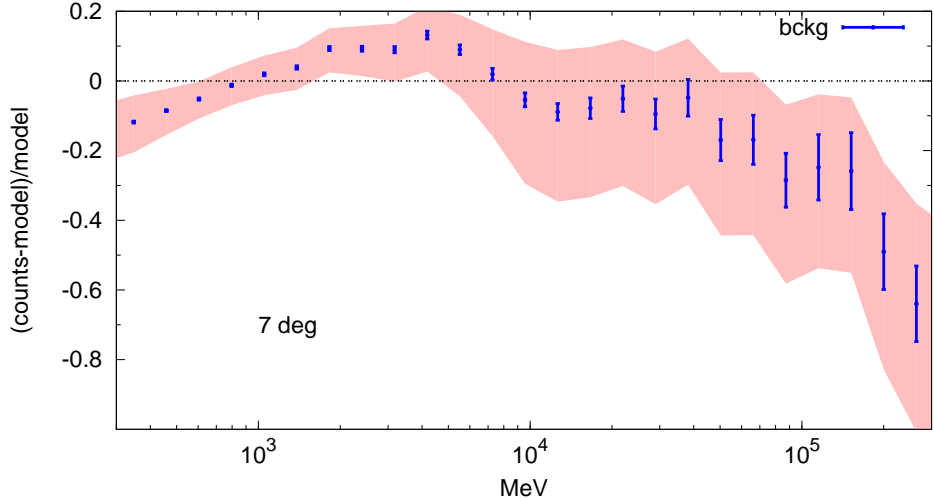,width=0.8\textwidth}
 \caption{\it
    The residuals in each bin of the available Fermi-LAT data~\protect\cite{fermi} used
     in Fig.~\ref{fig:latbckg},  for a 7 deg window around the GC
    in the range 300 MeV to 300 GeV,  as compared to the estimated background model.
    The shaded area represents the uncertainty in the effective area of the detector as discussed in the text.}
  \label{fig:residuals}
\end{figure}

In Fig.~\ref{fig:residuals} we  plot the residuals in each of the 25  bins in the range 300 MeV to 300 GeV,
i.e., the difference $\mathrm{(counts-model)/model}$, where in this case ``model" includes just the background sources. 
As discussed earlier,   the background models used to describe the collected data are able to reproduce them with an accuracy 
on the order of 25\%, especially for photon energies up to 100 GeV. This was expected, since the 
background models are not adequate  for energies larger that this~\cite{dgelat,rpslat}.
The shaded area represents the uncertainty in the effective area of the detector.
As discussed in~\cite{Vitale:2009hr},
the systematic uncertainty in the effective area of the LAT is currently estimated as 10\% at 100 MeV, 
decreasing to 5\% at 560 MeV, and then increasing to 20\% at 10 GeV. 
As suggested by in~\cite{Vitale:2009hr}, we assume that this uncertainty 
propagates to the model predictions.
With the systematic uncertainty included, we find $\chi^2 = 31.1$ for 25 degrees of freedom,
corresponding to a p-value of 19\%.

\section{Experimental Perspectives}
\label{expsect}

\begin{figure}[t!]
\epsfig{file=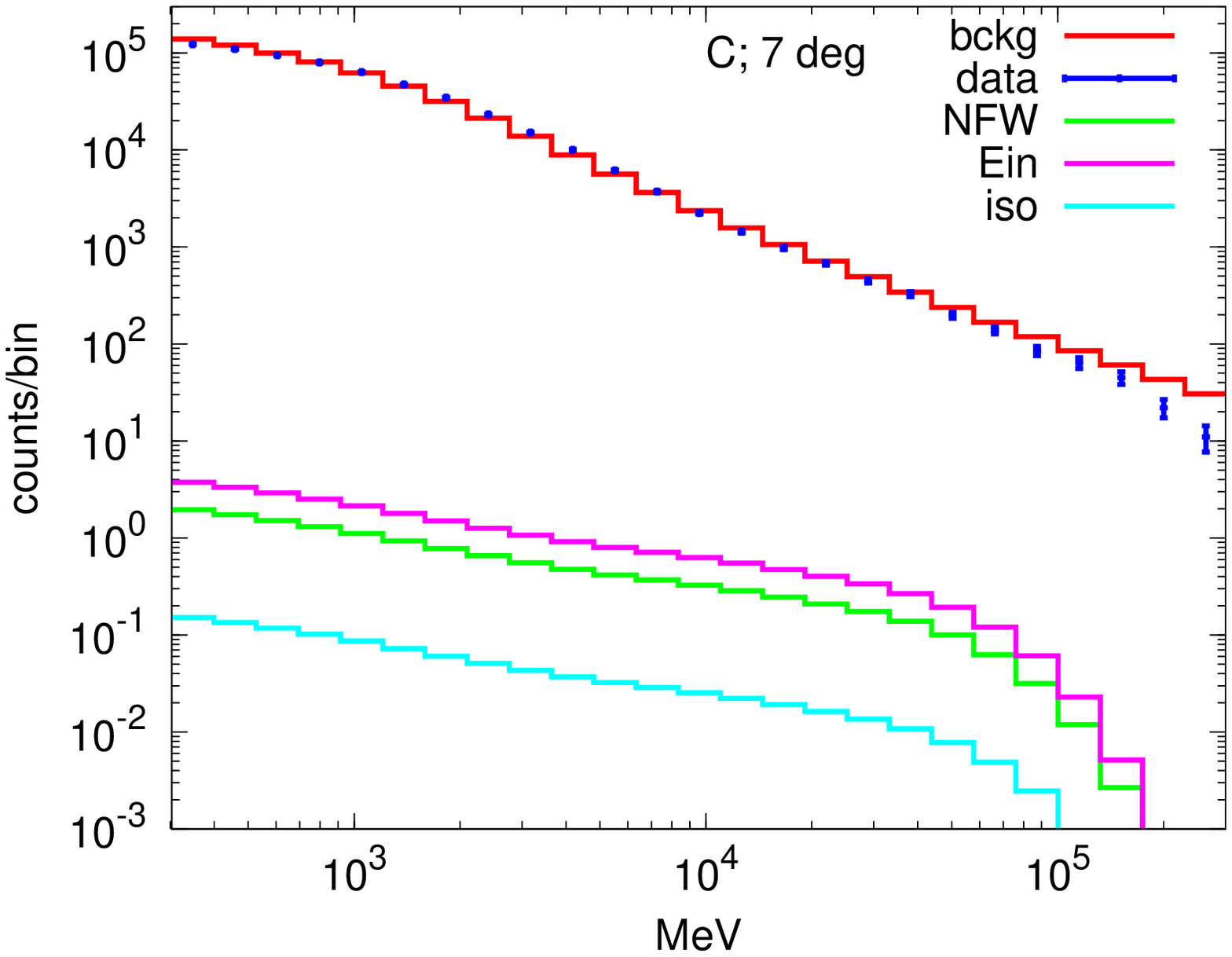,width=0.475\textwidth}
\epsfig{file=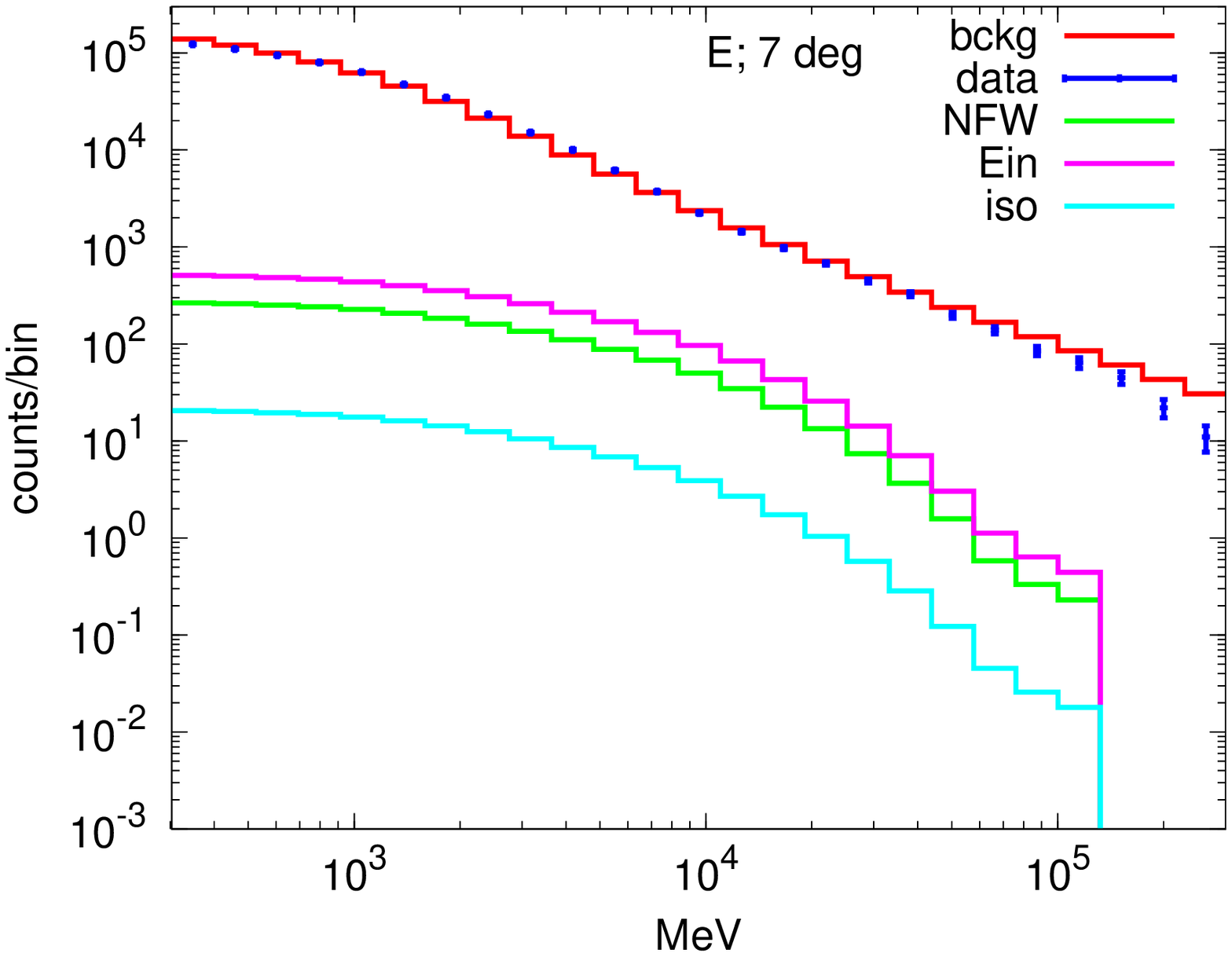,width=0.475\textwidth}\\
\epsfig{file=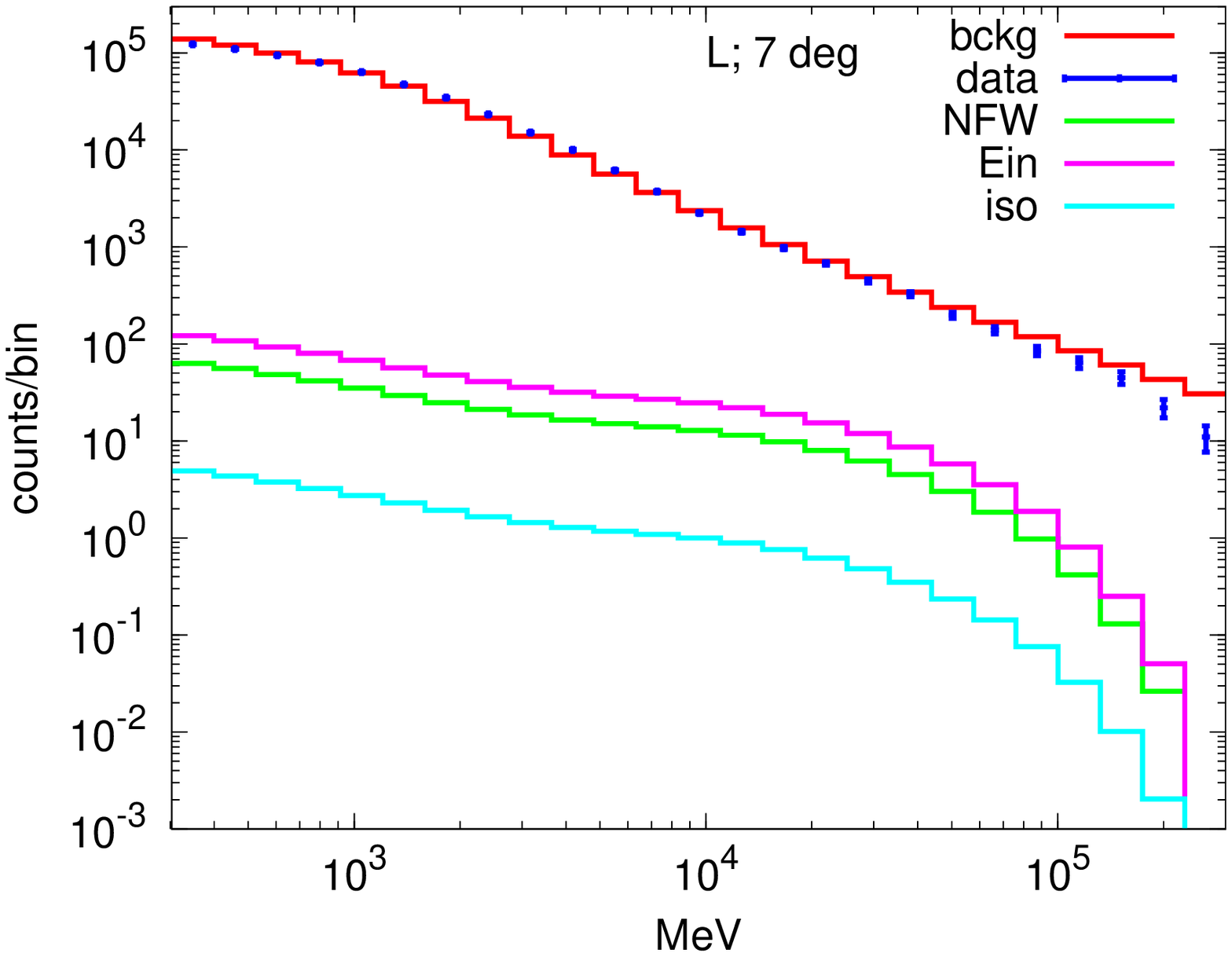,width=0.475\textwidth}
\epsfig{file=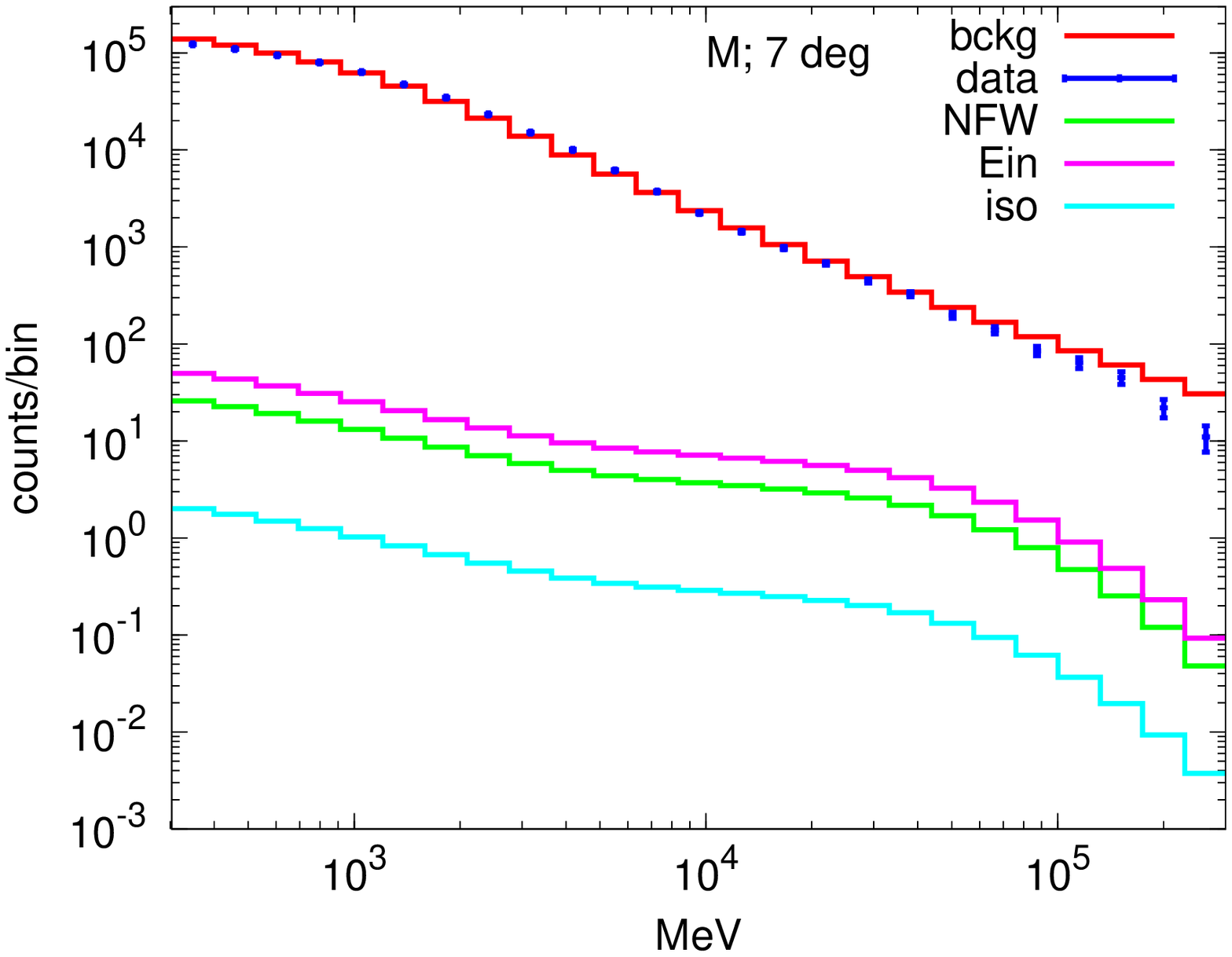,width=0.475\textwidth}

  \caption{\it
    The current Fermi-LAT data set~\protect\cite{fermi}, in counts per bin, as a function of photon 
    energy $E_\gamma$ in MeV, including the statistical error (blue points) for a 
    7-degree window around the GC. The red line is the corresponding estimated background. 
    We also display the signals expected for the three halo profiles we are using: NFW, Einasto and isothermal
    and the four
  panels correspond  to the four benchmark CMSSM points C,E,L and M, as in   Fig.~\ref{fig:fluxes}.  }
  \label{fig:counts}
\end{figure}

Using the results of the previous Section, we now compare the signals expected in CMSSM scenarios
with the background, under various hypotheses, again using 25 logarithmic bins in the range 300 MeV to 300 GeV.
We define the $\chi^2$ function as 
\beq
\chi^2=\sum_{i=1}^{25} \frac{(d_i-(b_i+s_i))^2}{\sigma_i^2} ,
\label{eq:chi2}
\eeq
where $d_i$ is the number of the data counts per bin,
$b_i$ is the expected background as calculated by Science Tools~\cite{ST}, and $s_i$ is the 
signal due to the dark matter halo for the corresponding bin. 
For the uncertainty, we have included both the statistical
uncertainty in the data count as well as the systematic uncertainty in the
effective area, $\sigma_{ea}$.  These have been added in quadrature, so that 
$\sigma_i^2 = d_i + \sigma_{ea}^2$.
In Fig.~\ref{fig:counts}, we present the current Fermi-LAT data set~\cite{fermi} as blue dots with statistical error bars, 
the background estimate (red curve) and with the signal corresponding to the 
three DM halo profiles discussed previously: NFW (green), Einasto (purple) and simple isothermal (cyan). 
The four panels of these figures correspond to the four CMSSM benchmark points C, E, L, M
discussed in the previous Section, so
the relative sizes of the signal can easily be understood. The coannihilation point C produces the weakest signal,
as expected, and the focus-point benchmark E gives the strongest signal. 
Points L and M yield signals that are almost comparable, again as
expected from Fig.~\ref{fig:fluxes}. Had we chosen a point from the right side of the funnel,
the signal would have been about a factor of 2 higher than shown for point M.

\begin{figure}[t!]
\begin{center}
\epsfig{file=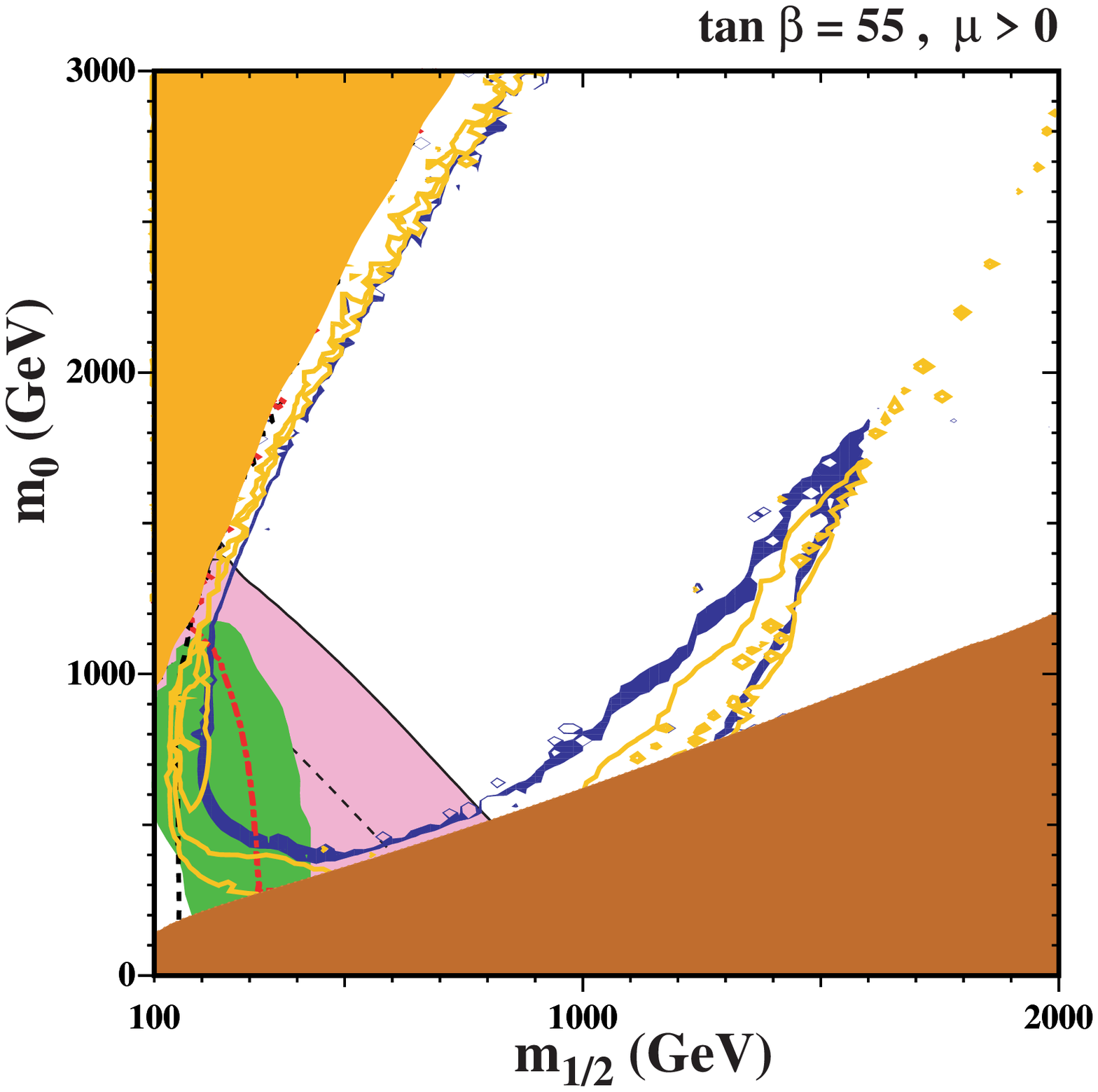,width=0.475\textwidth}
\epsfig{file=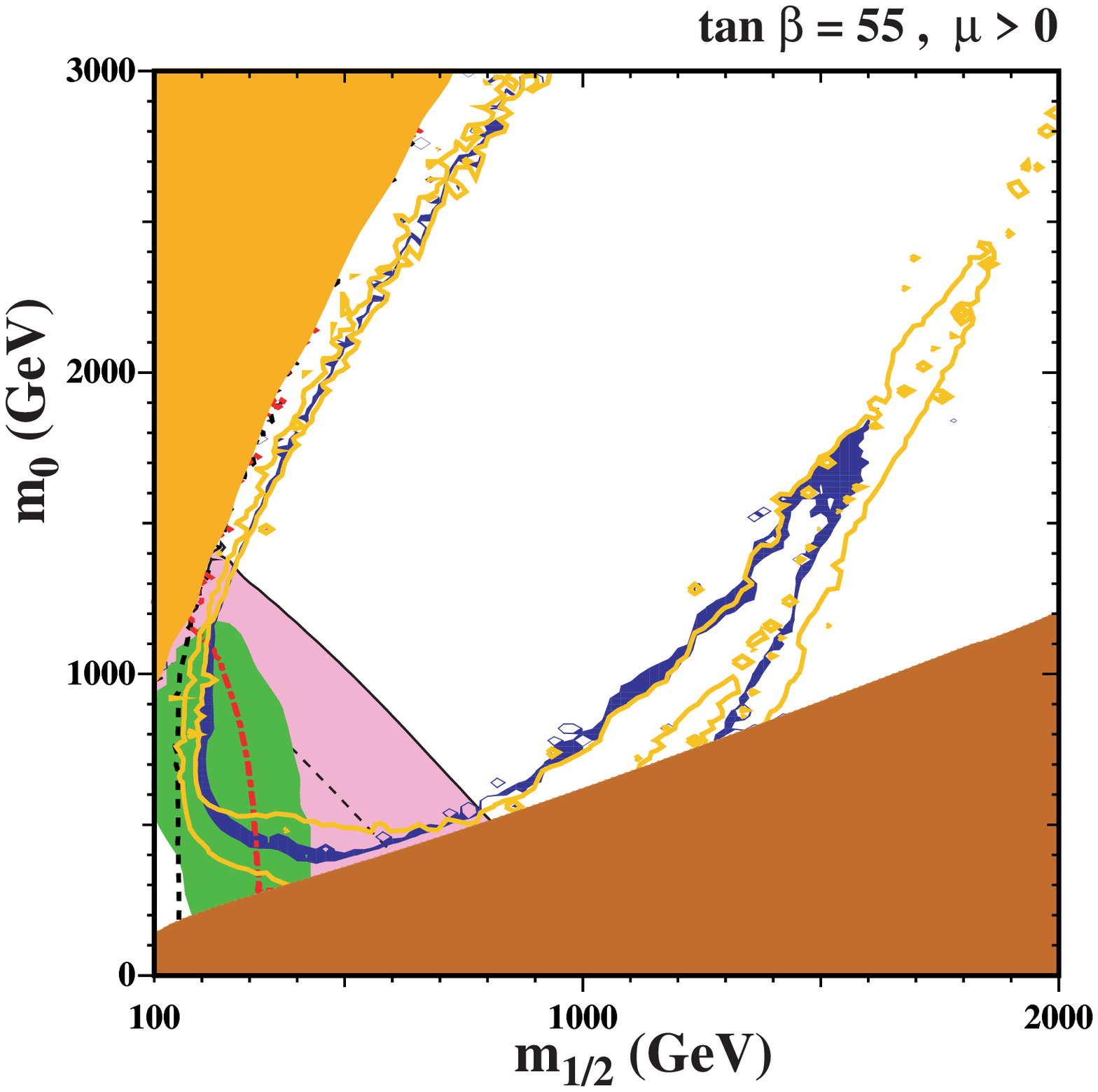,width=0.475\textwidth}\\
\caption{
\it Contours of  $\Delta \chi^2 = 0, 1, 4$ in the $m_{1/2},m_0$ plane for  $\tan \beta = 55$ (shown in yellow).
In the left panel, we assume an NFW profile for dark matter in the core of the galaxy, whereas in the right panel
we assume an Einasto profile. Conventions for the other curves and shaded regions are as for
Fig.~\ref{fig:planes}.}
\label{fig:tb55ne}
\end{center}
\end{figure}

We start with a nominal implementation
of the current experimental situation, that uses the present Fermi-LAT data set~\cite{fermi},
assumes the current understanding of the background and systematic errors, and uses an
NFW density profile.
As a preliminary to examining the signals along the WMAP strips
shown in Fig.~\ref{fig:planes}, we first consider a more global view across 
an $(m_{1/2},m_0)$ plane in the CMSSM.  As described in the previous section, 
in the absence of any supersymmetric contribution, our modeling of the astrophysical
background give a value of $\chi^2 = 31.1$ for 25 degrees of freedom.
In Fig.~\ref{fig:tb55ne}, we show for illustration (in yellow) contours of constant $\Delta \chi^2$ across the CMSSM
$(m_{1/2},m_0)$ plane for $\tan \beta = 55$~\footnote{We do not show analogous results for 
$\tan \beta = 10$, since the results are unpromising, as we see below when we restrict our attention to the WMAP strips.},
 exhibiting contours of $\Delta \chi^2 = 0,1$ and 4
relative to the value of $\chi^2$ without supersymmetry~%
\footnote{We recall that the relic LSP density is too large in the interior of the plane
between the WMAP strips, and too small in the outer regions between the WMAP strips and the charged-LSP and
electroweak symmetry breaking boundaries. The interior region is forbidden in the $R$-conserving
scenario discussed here, whereas the outer regions could be acceptable in the presence of some other
component of dark matter. However, in this case the annihilation rate should be reduced by the square
of the relic LSP density relative to the WMAP dark matter density. Only the intersections of the $\Delta \chi^2$
contours with the WMAP strips should be taken literally.}.
Note that there is a small region at low $m_{1/2}$, in the region already excluded by 
$b \to s \gamma$, where there is a small but insignificant improvement to $\chi^2$.
As one can see in the left panel of Fig.~\ref{fig:tb55ne}, where an NFW profile has been assumed,
a modest constraint is imposed by the present Fermi data.  In addition to 
the area at small $m_{1/2}$ that is already excluded by supersymmetry searches at 
LEP, the $\Delta \chi^2 = 1$ contour runs between the branches of the rapid-annihilation funnel, and
disfavours the right branch at the 68\% CL.  The $\Delta \chi^2 = 4$ contour runs deep inside the funnel
where the relic density is below the WMAP range.
In the right panel, where the Einasto profile has been assumed, the disfavoured region is somewhat 
larger, as expected from the values of the halo factors in the Table~\ref{table:halo}. 
On the other hand, based on these values,  the opposite is expected for the 
for the isothermal model.

\begin{figure}[th!]
\begin{center}
\epsfig{file=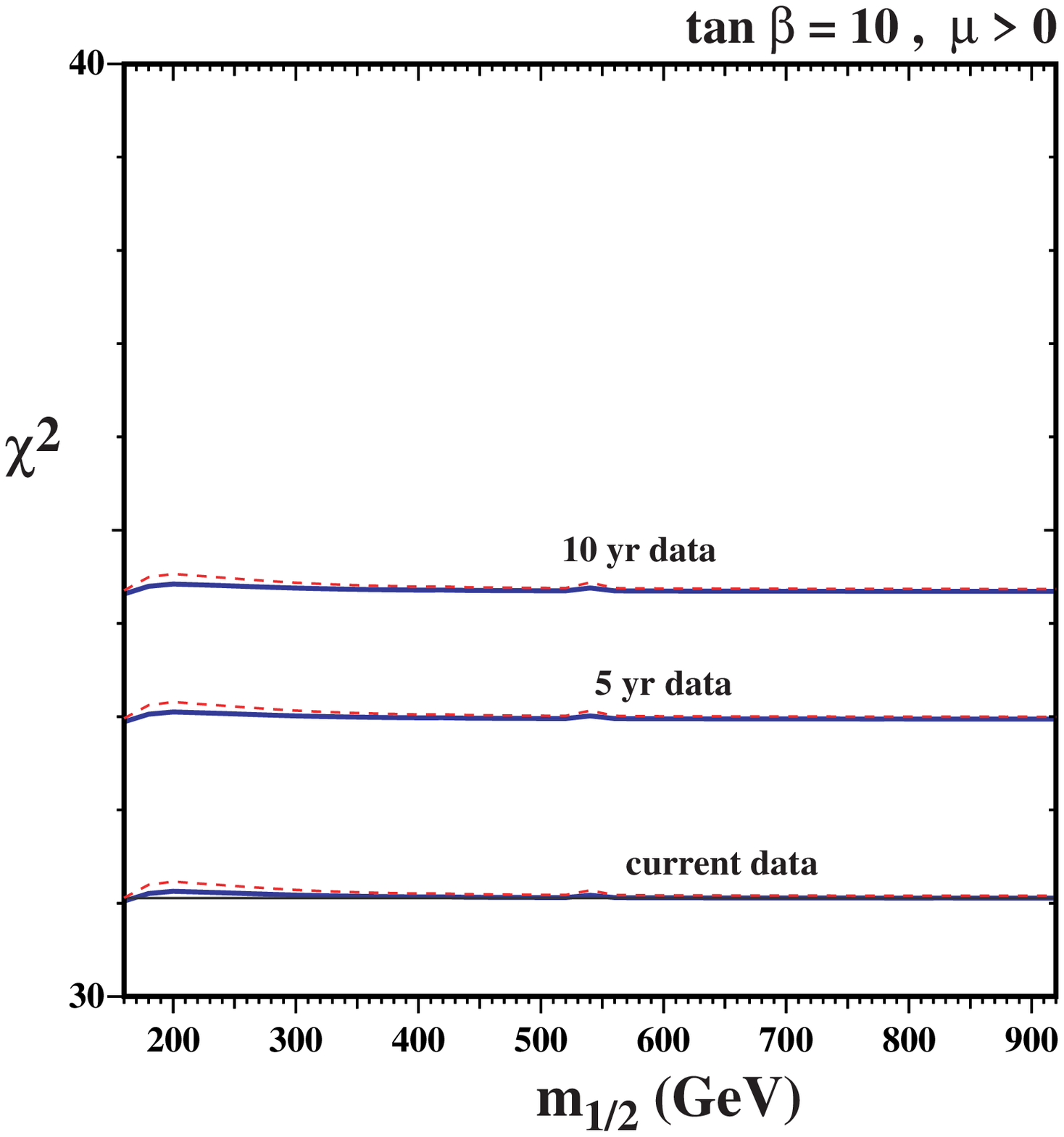,width=0.46\textwidth}
\epsfig{file=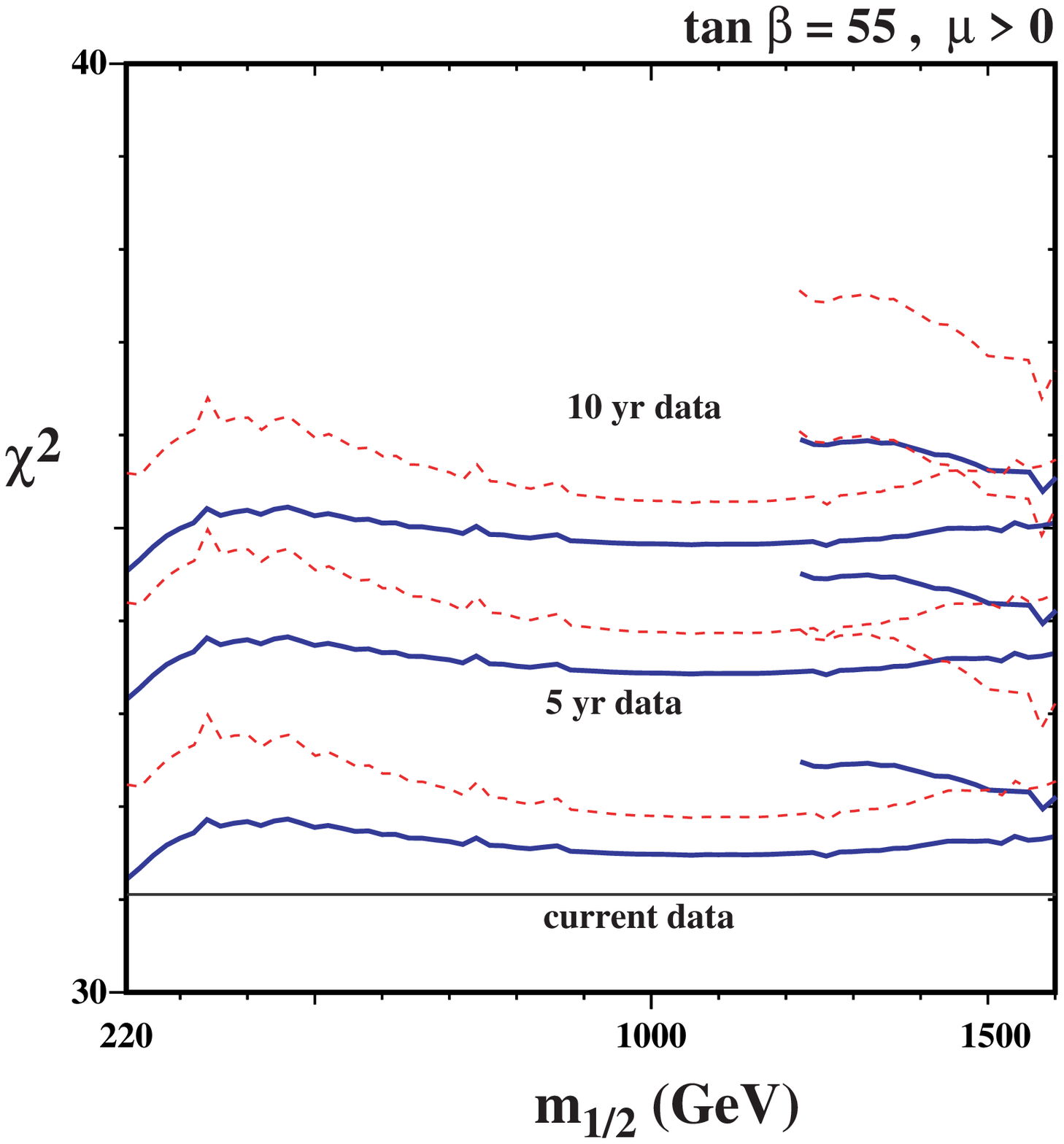,width=0.46\textwidth}\\
\epsfig{file=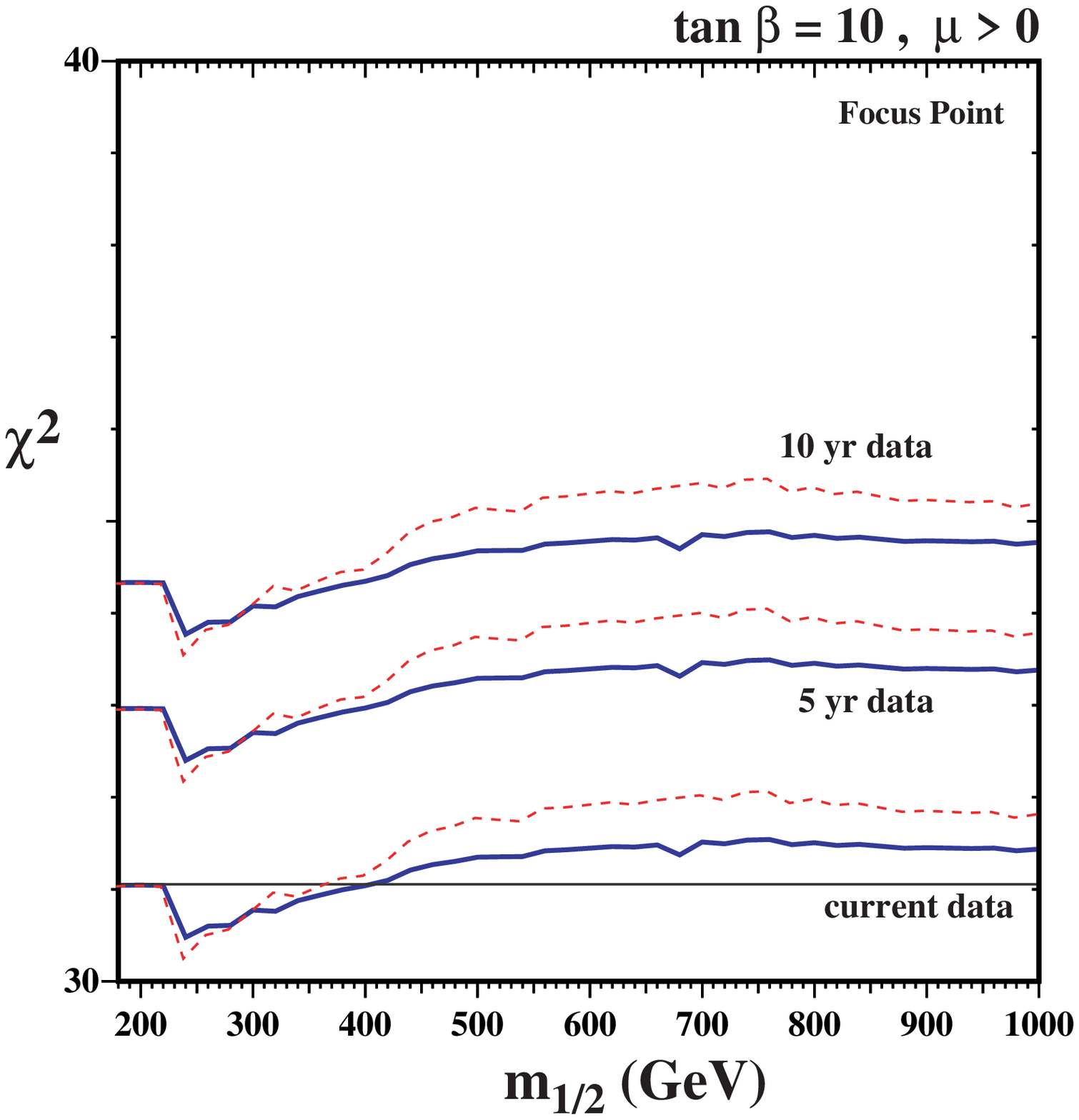,width=0.46\textwidth}
\epsfig{file=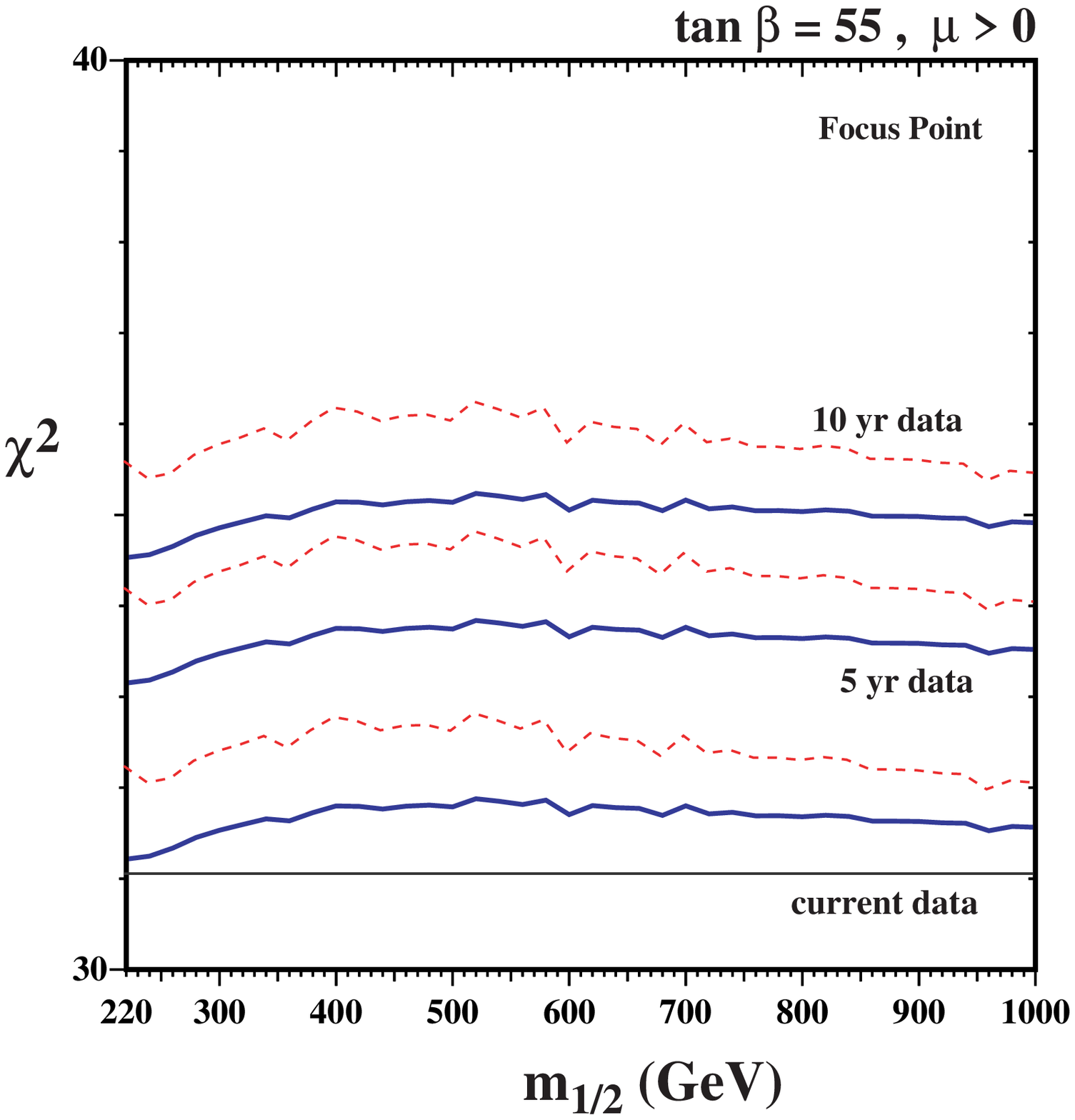,width=0.46\textwidth}
\caption{
\it The $\chi^2$ functions along the
CMSSM WMAP strips as functions of $m_{1/2}$  for $\tan \beta = 10$
(left panels) and $\tan \beta = 55$ (right panels), in the coannihilation
and funnel regions (upper panels) and in the focus-point region
(lower panels). In each panel, we display the $\chi^2$ function
for the background alone, a horizontal line at $\chi^2 = 31.1$, and
the $\chi^2$ function obtained by adding the calculated $\chi - \chi$
annihilation signal in the current (approximately 2 1/2 year) Fermi data sample and
in projected 5- and 10-year data sets. Solid (blue) curves are based on an NFW profile,
while dashed (red) curves are based on an Einasto profile.}
\label{fig:nominal}
\end{center}
\end{figure}

We now focus on the WMAP strips for $\tan \beta = 10$ and 55.
The results of our nominal implementation of the current and
possible future Fermi constraints are shown in Fig.~\ref{fig:nominal}
for the coannihilation and funnel strips (upper panels) and $\tan \beta = 10$
(left panels) and $\tan \beta = 55$ (right panels). As a function of $m_{1/2}$
along each of these lines, we display the $\chi^2$ function for the background alone,
which is a constant 31.1 for 25 degrees of freedom, and the
corresponding $\chi^2$ function for the combination of the background with the
signal calculated as a function of $m_{1/2}$ in each scenario, for the current
Fermi-LAT data~\cite{fermi}. Also shown are the possible results from future 5- and 10-year data sets. 
The baseline $\chi^2$ is correspondingly higher due to the expected improvement
in the statistical uncertainty in the data, assuming no improvement 
in the systematic uncertainty associated with the effective area. 
The horizontal lines assuming no supersymmetry are not shown
for these hypothetical data sets. In each case, the results assuming an NFW
profile are shown as solid lines, and results assuming an Einasto profile are
shown as dashed lines.

We see immediately that the calculated signal with the current Fermi-LAT data set~\cite{fermi}
has very little effect on the overall $\chi^2$ function along the
coannihilation strip for $\tan \beta = 10$ (upper left panel), whereas somewhat larger
effects are visible along the WMAP strips shown in the other panels.
However, in all cases except along the second funnel strip for $\tan \beta = 55$
the change in the $\chi^2$ function is $< 1$ in the case of the NFW profile (solid blue lines),
and hence not significant. In general, adding the supersymmetric signal worsens 
slightly the quality of the fit, though there is a slight improvement along the
focus-point strip for $\tan \beta = 10$ for $200~{\rm GeV} < m_{1/2} < 400$~GeV.
The (red) dashed curves in Fig.~\ref{fig:nominal} show the resulting 
$\chi^2$ functions assuming an Einasto dark matter profile.
While this does yield somewhat larger effects,  we
see that $\Delta \chi^2 \la 3$ even along the right branch of the funnel for $\tan \beta = 55$, and 
a 95\% CL exclusion is not possible.

\begin{figure}[t!]
\begin{center}
\epsfig{file=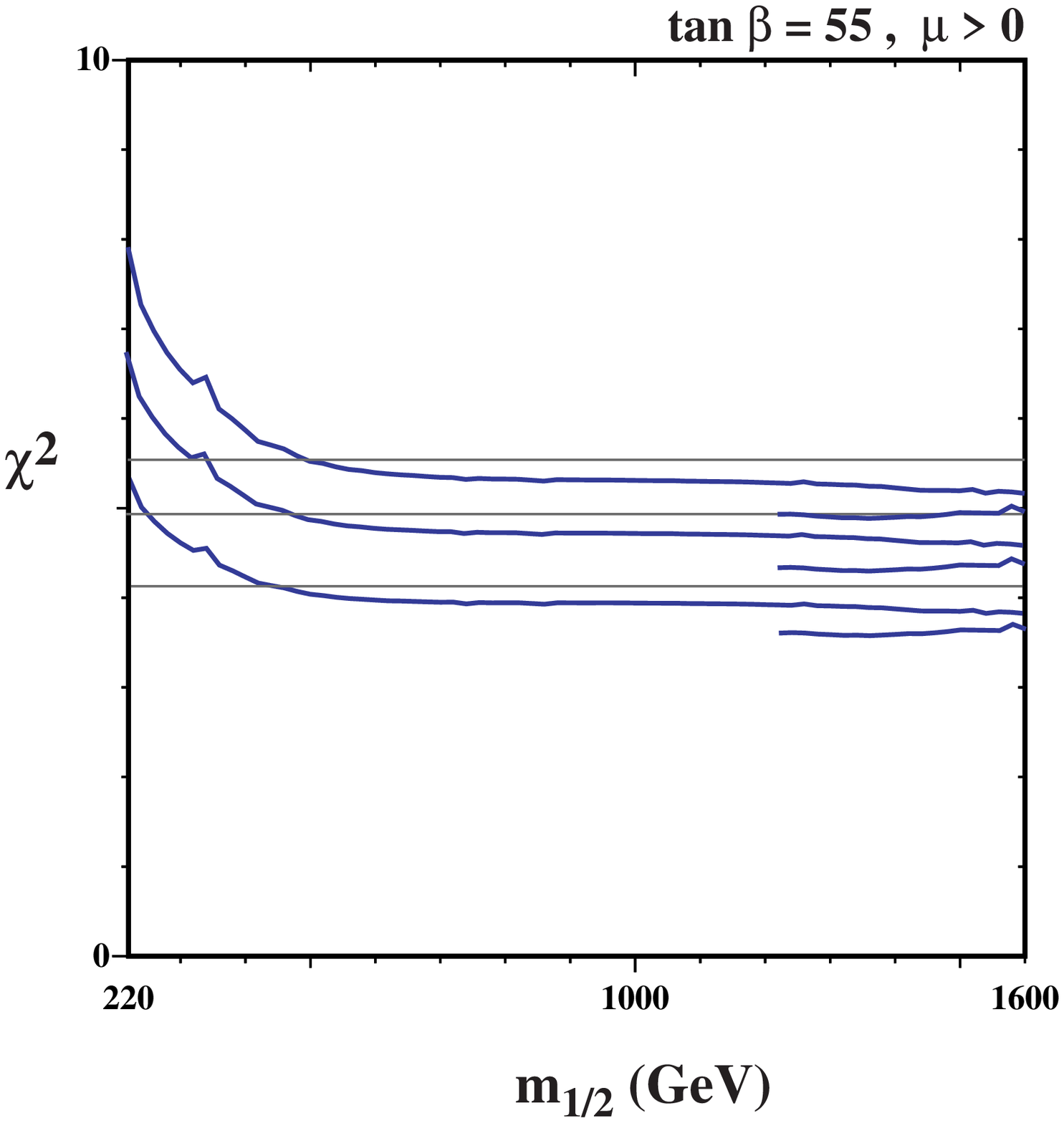,width=0.475\textwidth}
\epsfig{file=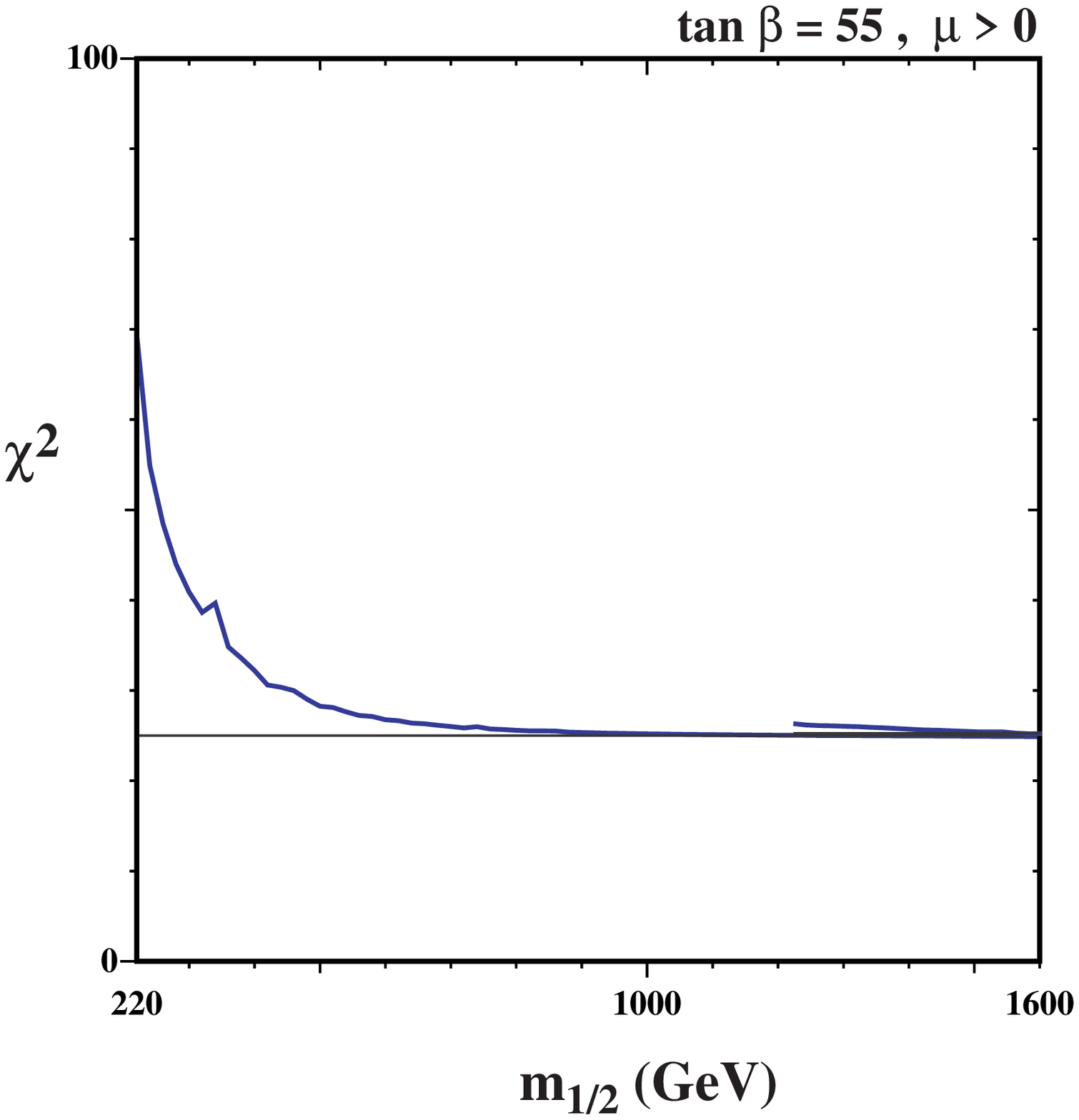,width=0.475\textwidth}\\
\epsfig{file=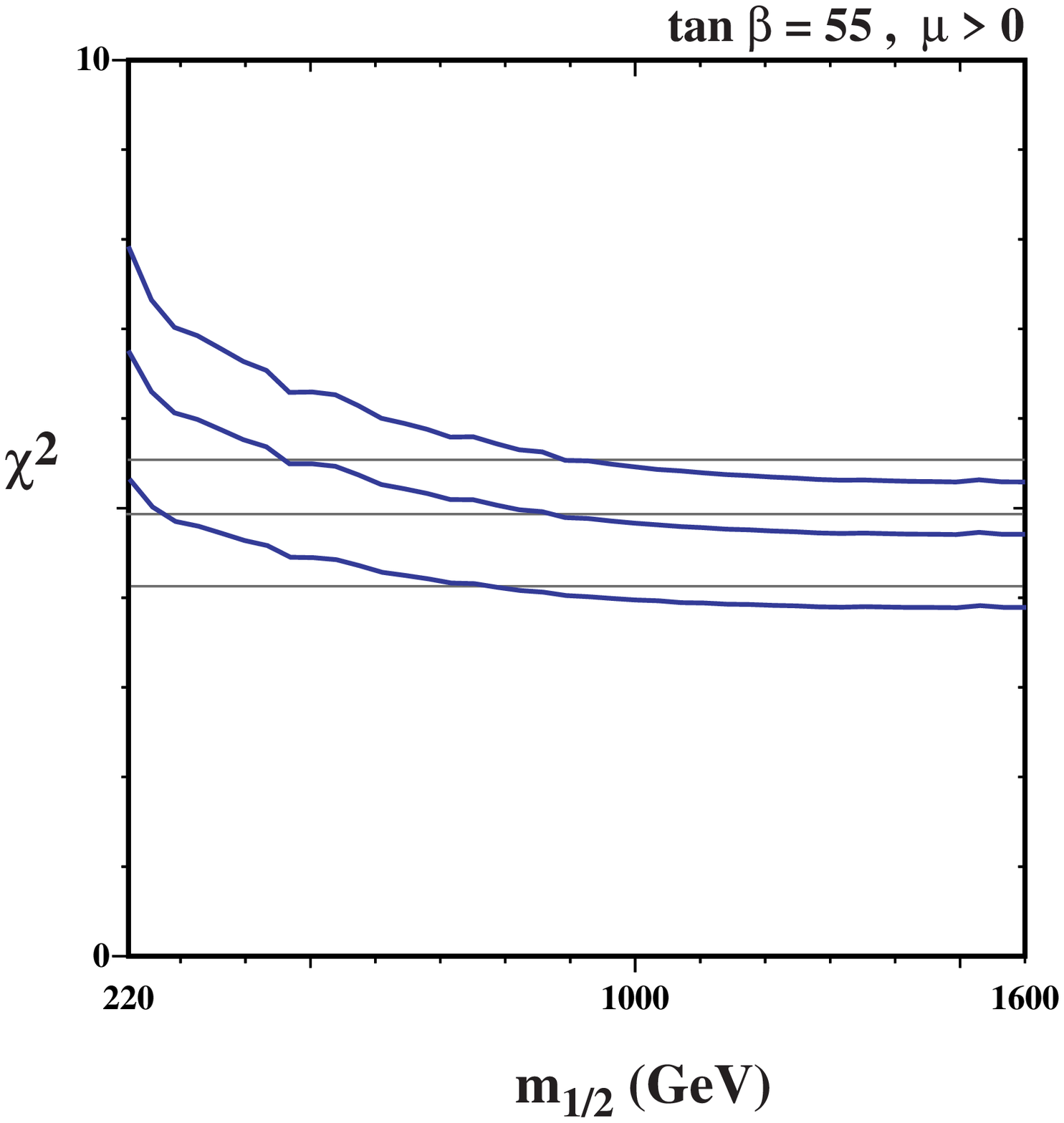,width=0.475\textwidth}
\epsfig{file=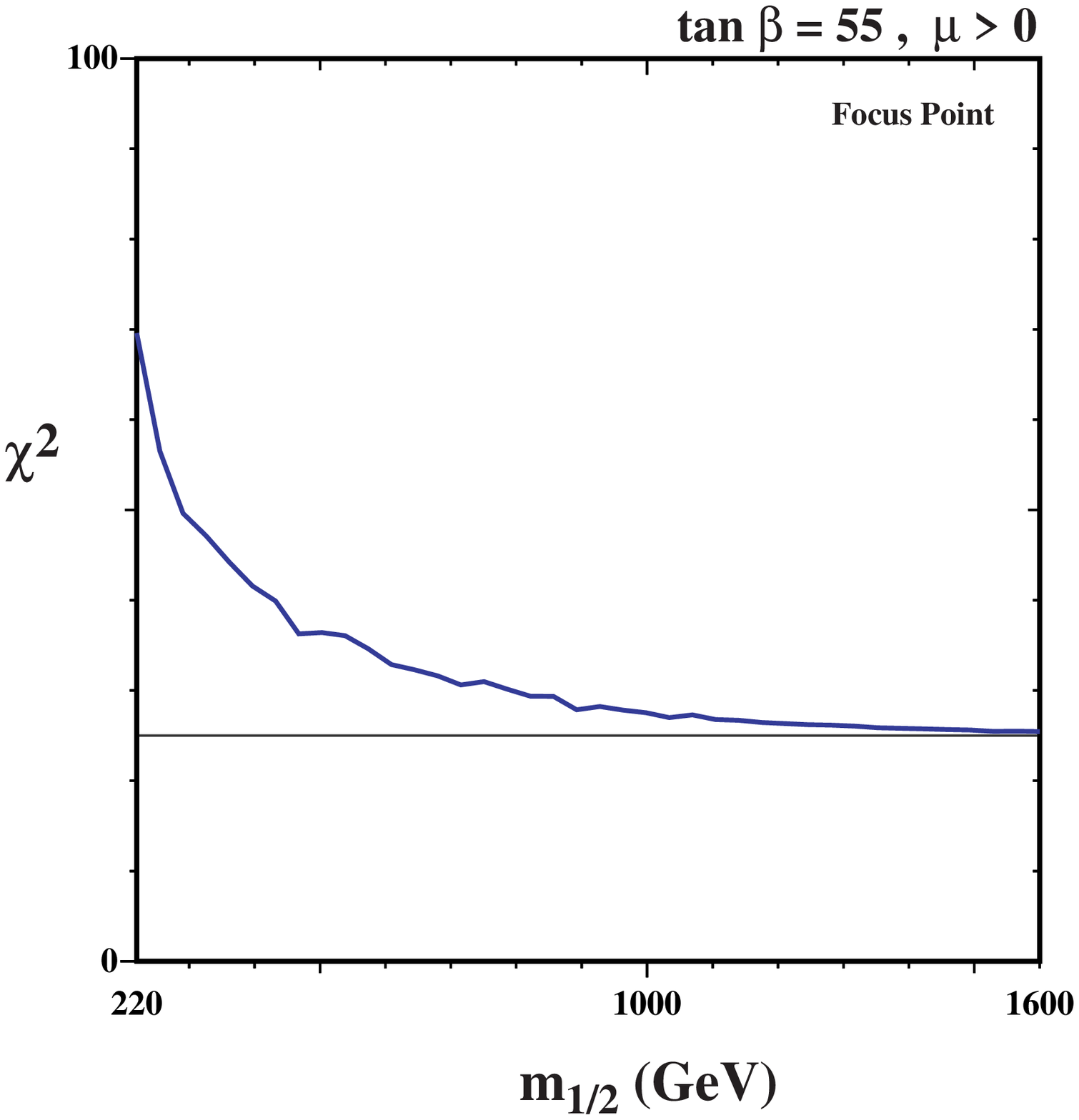,width=0.475\textwidth}
\caption{
\it The $\chi^2$ functions as functions of $m_{1/2}$ along the
CMSSM focus-point WMAP strip for $\tan \beta = 55$ assuming a factor of
2, 3 or 4 reduction in the systematic error  $\sigma_{ea}^2$ (left panels) or a negligible
systematic error (right panels), assuming in all cases that an improved
estimate of the background brings it to $\pm 1 \sigma$ from the data. 
The upper (lower) panels are for the coannihilation/funnel strip (focus-point strip).}
\label{fig:improved}
\end{center}
\end{figure}

\begin{figure}[t!]
\begin{center}
\epsfig{file=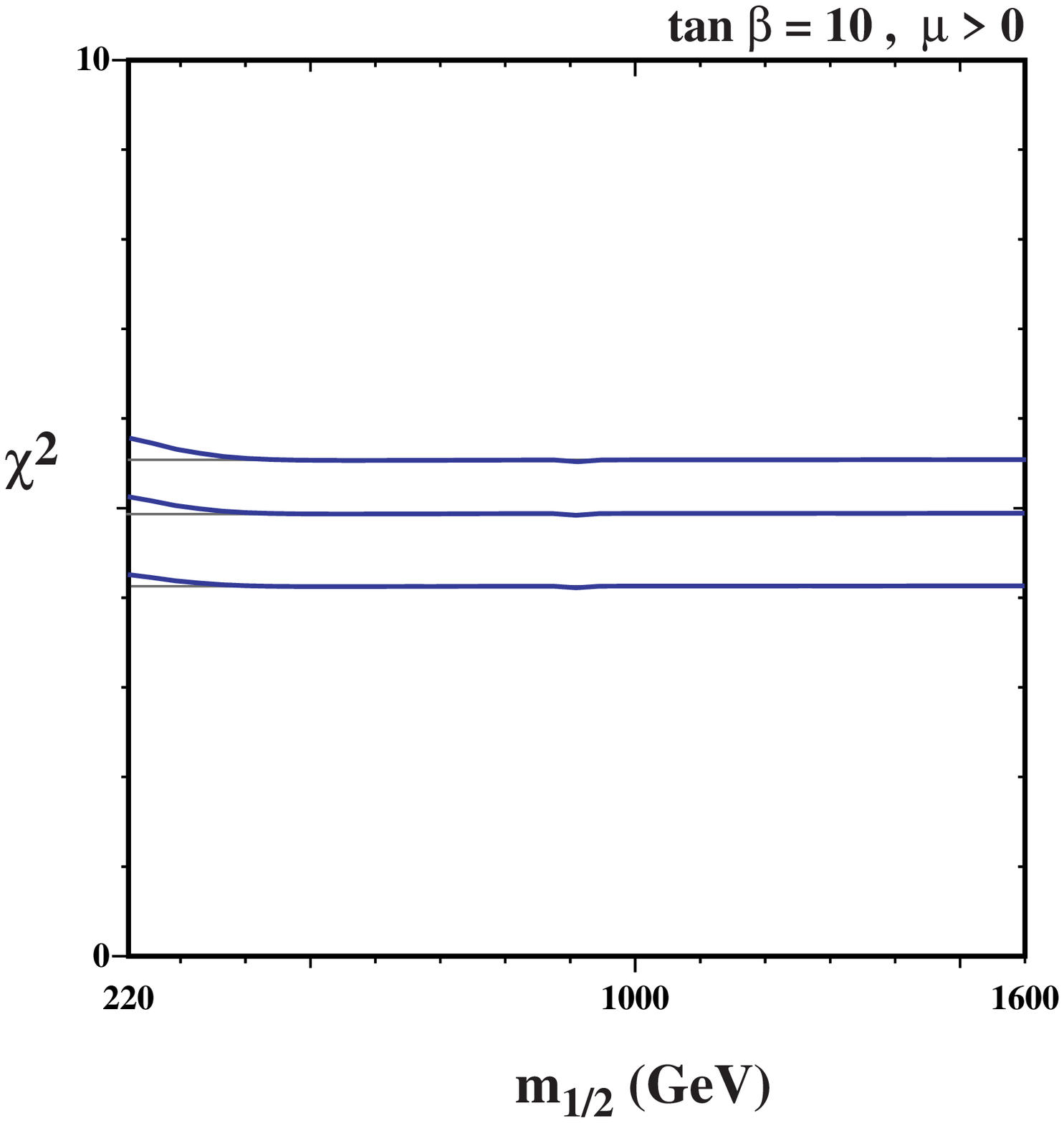,width=0.475\textwidth}
\epsfig{file=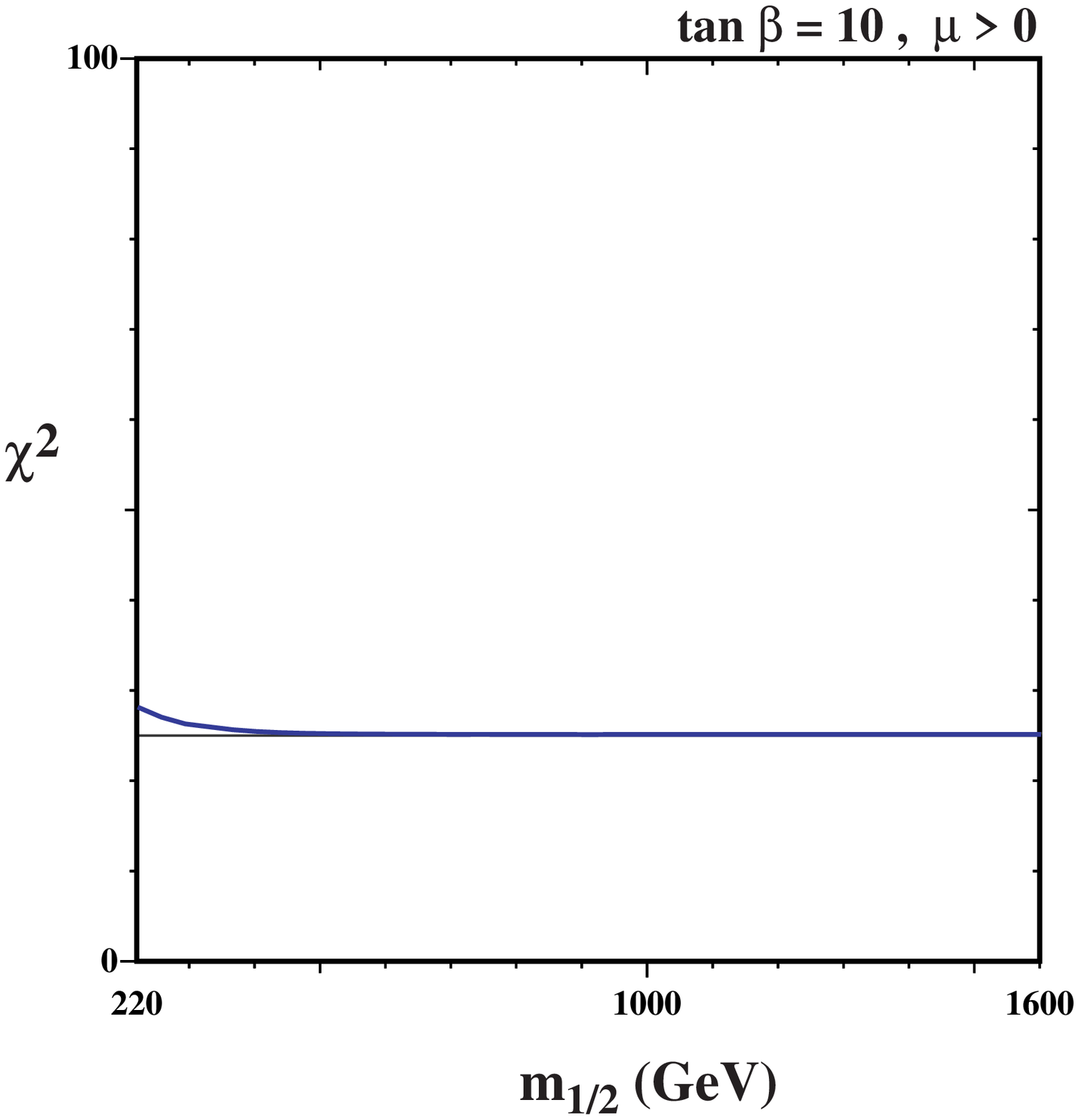,width=0.475\textwidth}\\
\epsfig{file=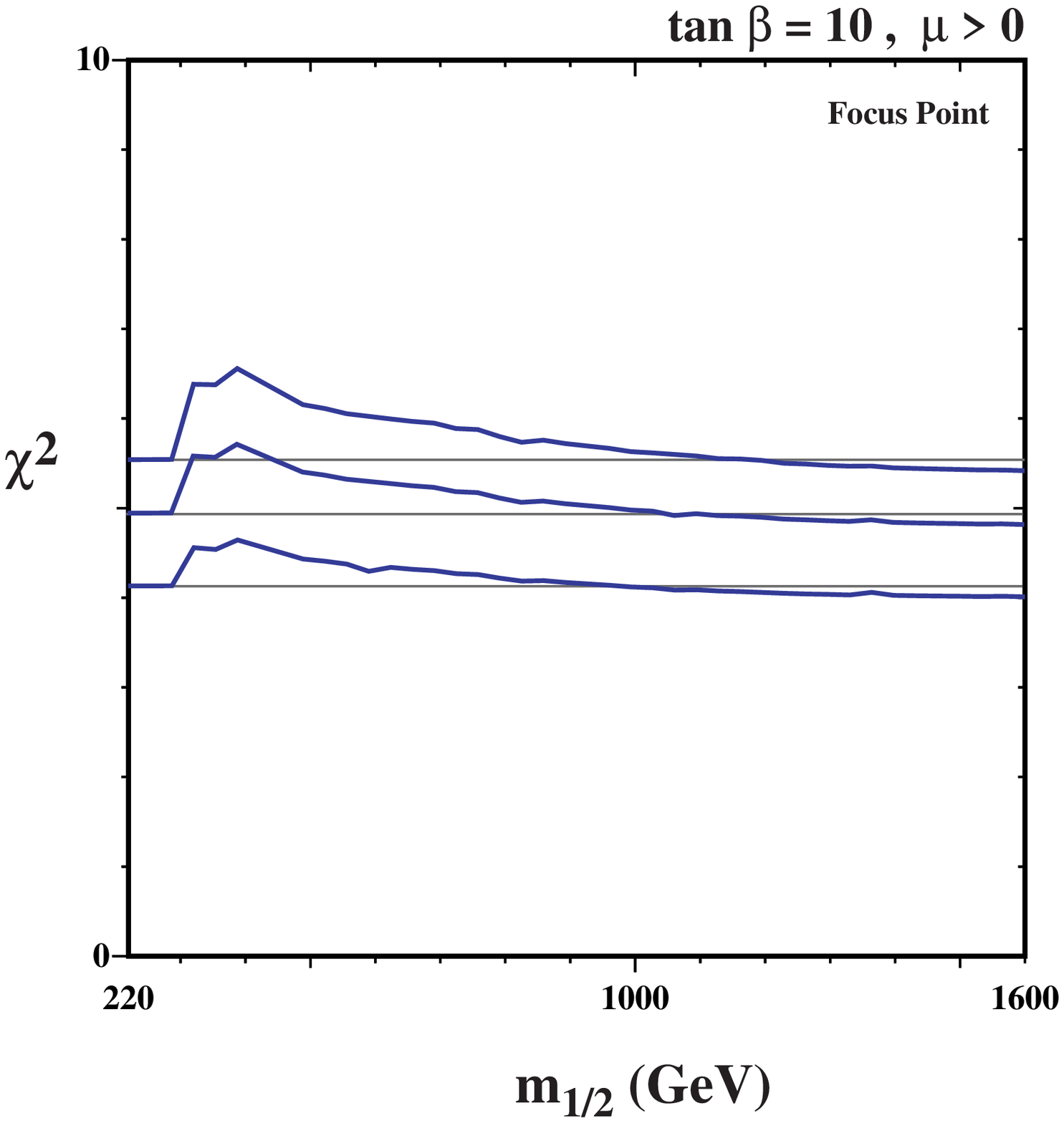,width=0.475\textwidth}
\epsfig{file=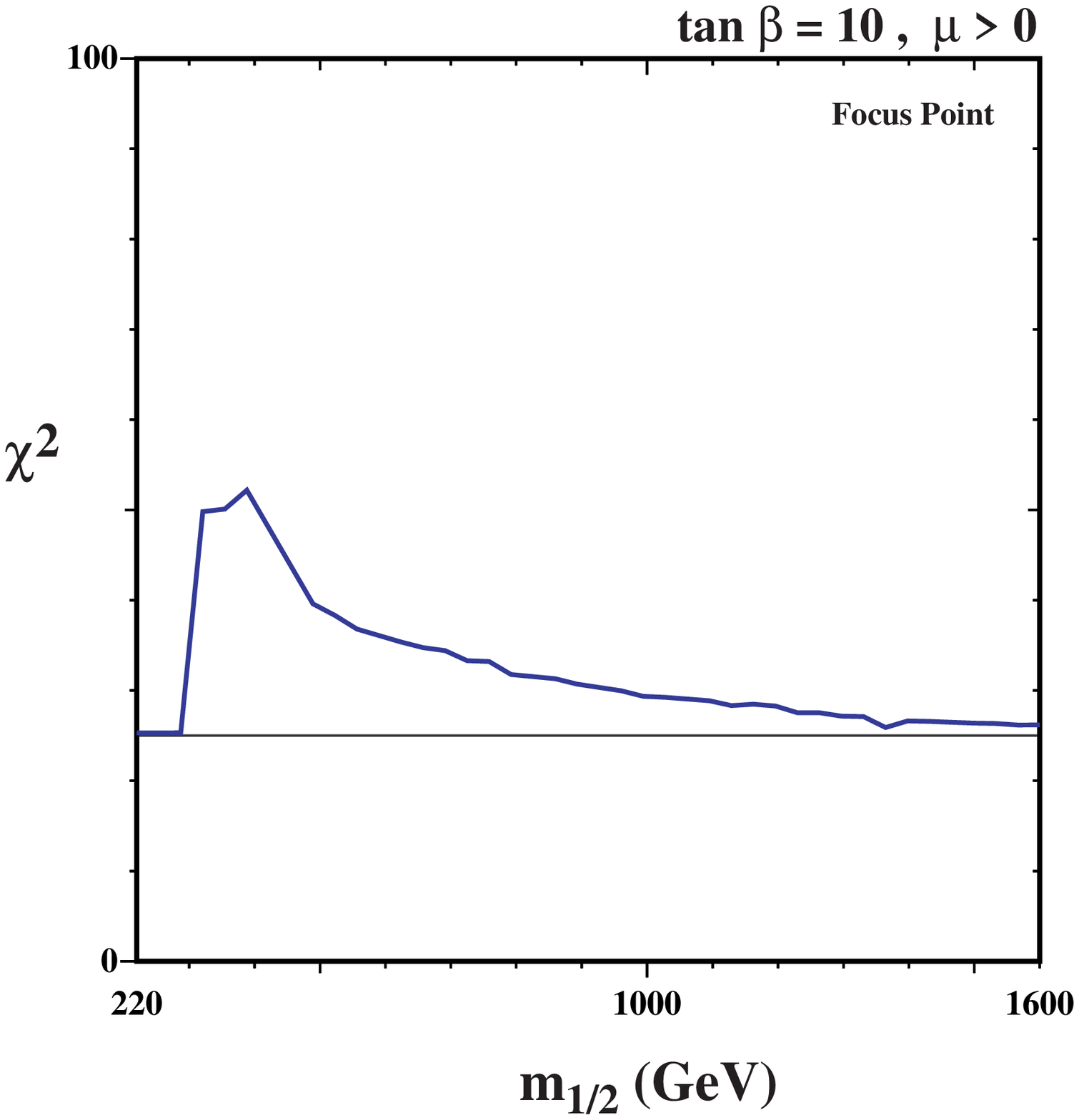,width=0.475\textwidth}
\caption{
\it As in Fig.~\protect\ref{fig:improved}, but for $\tan \beta = 10$.}
\label{fig:improved10}
\end{center}
\end{figure}

Hence, the most that one could hope for along most of the WMAP strips studied
is some inconclusive level of disfavour, except in the limited region 
of small $m_{1/2}$ for $\tan \beta = 10$ where some
improvement in the overall $\chi^2$ of a global fit might eventually be possible.
Significant improvements in our
current understandings of the background and/or systematic
errors would be required
before the scenarios studied could be constrained significantly.

Examples of the effects of possible improvements in understanding the
background and systematic errors are shown 
in Fig.~\ref{fig:improved}, in the cases of the coannihilation/funnel strip (upper panels)
and  focus-point strip (lower panels) WMAP strips with $\tan \beta = 55$,
assuming an NFW profile. The $\chi^2$ functions in the absence of any supersymmetric
signal are shown as horizontal black lines, and the effects of including a signal are
shown as blue lines. In all of these panels,
we have assumed that improvements to our understanding of the background 
bring it to lie within 1 $\sigma$ of the data, with the sign of the discrepancy chosen
randomly in each bin. In the left panels, we also assume reductions
in the current systematic error in the effective area by a factors of 2, 3 and 4 in $\sigma_{ea}^2$, whereas
in the right panels the systematic error is assumed to be negligible~\footnote{For large $m_\chi$ 
even a small signal improves slightly the fit, in comparison to the background case only. So we see 
the $\chi^2$ of the signal is smaller than the $\chi^2$ of the background for large $m_{1/2}$.
This improvement, for the case of the reduced systematic error, is due to the $\chi^2$ contribution of the last 
bins that contain energetic photons with  $E_\gamma> 150 \gev$.}. 
We see immediately in the left panels that reducing the systematic
error by a factor of 2 does not improve greatly the discrimination between
the background-only and signal + background hypotheses. Even a reduction
by a factor of 4 does not lead to  a 95\% CL exclusion limit, whereas we see
in the right panels that a reduction in the systematic error to a negligible level might 
provide such discrimination. Depending on the evolution of the understanding of the
background, adding a CMSSM might affect
significantly the fit for small $m_{1/2}$.
With the present background estimate, 5 or 10 years of Fermi data
might provide evidence for the CMSSM at small $m_{1/2}$, albeit in the region
disfavoured by $b \to s \gamma$, or possibly evidence against the CMSSM in the
region $m_{1/2} > 500$~GeV where accelerator constraints are currently
irrelevant. The corresponding figures for $\tan \beta = 10$ with improved systematics are shown in 
Fig.~\ref{fig:improved10}.  In the case of the coannihilation strip for $\tan \beta = 10$, even the absence
of systematic errors would not allow significant information to be obtained, though some
information might be obtainable about the focus-point strip.
For instance, using the Einasto profile instead of NFW, 
the $\chi^2$ of the last panel  in the Fig.~\ref{fig:improved10}    is about 3 times bigger,
 enabling  more definite conclusions to be drawn.

\section{Summary and Prospects}

We have studied in this paper the potential of searches by the Fermi-LAT detector~\cite{fermi} for
$\gamma$ rays from the Galactic centre produced by the annihilations
of neutralino LSPs in the CMSSM. We have shown predictions for benchmark points,
$(m_{1/2}, m_0)$ planes, and the concentrated on the WMAP strips
for $\tan \beta = 10$ and 55. We have found that the $\gamma$-ray
signal would be very difficult to detect along the coannihilation strip
for $\tan \beta = 10$, as expected from the relatively small $\chi \chi$
annihilation cross section along this strip. The focus-point strip for
$\tan \beta = 10$ and both the coannihilation/funnel and
focus-point strips for $\tan \beta = 55$ have larger
annihilation cross sections and offer better prospects, though even
in these cases the present data do not offer good prospects for
discriminating between models, because of uncertainties in
the background estimates and the current systematic errors of
the Fermi-LAT detector. 

Future Fermi-LAT data sets might offer some information about parameters
of the CMSSM, also in regions of parameter space not accessible to
accelerator experiments. However, this would require several more years of data 
and a substantial reduction in the present systematic error.
This is therefore a priority for optimizing the prospects for future Fermi-LAT
constraints on the CMSSM from searches for $\gamma$ rays from the 
Galactic centre. Quantitative understanding of the constraints imposed
by Fermi-LAT results would also require advances in our understanding of
the dark matter density profile in the core of the Galaxy, since we see
considerable differences between results assuming NFW and Einasto profiles.

The Fermi-LAT detector also has data on $\gamma$ rays from many
other astrophysical sources, including the Galactic bulge, dwarf
galaxies and diffuse emissions. We therefore plan to extend our
analysis to these cases. However, it is clear that it will always be 
more difficult to pin down the CMSSM than some other supersymmetric models
via searches for energetic $\gamma$ rays from astrophysical sources,
and the same remark applies to other astrophysical strategies to look for neutralino
annihilations.

\vspace*{1cm}
\section*{Acknowledgements}

The work of J.E. was supported partly by the London
Centre for Terauniverse Studies (LCTS), using funding from the European
Research Council 
via the Advanced Investigator Grant 267352. 
The work of K.A.O. was supported in part
by DOE grant DE--FG02--94ER--40823 at the University of Minnesota.
K.A.O. also thanks SLAC 
(supported by the DOE under contract number DE-AC02-76SF00515) and 
the Stanford Institute for Theoretical Physics
for their hospitality and support. The work of V.C.S. was supported by Marie Curie International
  Reintegration grant SUSYDM-PHEN, MIRG-CT-2007-203189.
 We acknowledge contributions by Yudi Santoso in the early stages of this work.
  We thank Elliott Bloom, Aldo Morselli,
Simona Murgia and Stefano Profumo for helpful discussions about Fermi-LAT data.
 V.C.S. thanks Johann Cohen-Tanugi, Marco Cirelli and  Catena Riccardo  for 
 useful discussions.

\end{document}